\documentclass[3p]{elsarticle} 

\usepackage{lineno,hyperref}

\usepackage{amsmath,amssymb}
\usepackage{todonotes}
\usepackage{caption}
\usepackage{subcaption}
\usepackage{graphicx}
\usepackage{pgfplots}
\usepackage{siunitx}
\DeclareUnicodeCharacter{2212}{-}
\usepackage{import,bm,float}
\newcommand{\defeq}{\stackrel{\text{def}}{=}}

\journal{}

\newcommand{\boldface}[1]{\boldsymbol{#1}}  
\newcommand{\bfa}{\boldface{a}}

\newcommand{\bfe}{\boldface{e}}
\newcommand{\bff}{\boldface{f}}
\newcommand{\bfg}{\boldface{g}}

\newcommand{\bfk}{\boldface{k}}

\newcommand{\bfu}{\boldface{u}}

\newcommand{\bfx}{\boldface{x}}

\newcommand{\bfA}{\boldface{A}}
\newcommand{\bfB}{\boldface{B}}

\newcommand{\calL}{\mathcal{L}}

\newcommand{\calR}{\mathbb{R}}
\newcommand{\calS}{\mathcal{S}}

\newcommand{\T}{^{\mathrm{T}}} 

\newcommand{\Rset}{\mathbb{R}}

\newlength{\boxwidth}
\def\dd{\;\!\mathrm{d}}

\def\btheorem{\begin{theorem}}
\def\etheorem{\end{theorem}}
\def\blemma{\begin{lemma}}
\def\elemma{\end{lemma}}
\def\bproposition{\begin{proposition}}
\def\eproposition{\end{proposition}}
\def\bcorollary{\begin{corollary}}
\def\ecorollary{\end{corollary}}
\def\bdefinition{\begin{definition}}
\def\edefinition{\end{definition}}
\def\bexample{\begin{example}}
\def\eexample{\end{example}}
\def\bremark{\begin{remark}}
\def\eremark{\end{remark}}

\DeclareMathOperator{\divv}{div}

\DeclareMathOperator{\argmin}{{arg\,min}}

\newcommand{\be}{\begin{equation}}
\newcommand{\ee}{\end{equation}}

\newcommand{\beq}{\begin{eqnarray*}}
\newcommand{\eeq}{\end{eqnarray*}}
\newcommand{\bem}{\begin{multline}}
\newcommand{\eem}{\end{multline}}
\newcommand{\ba}{\begin{align*}}
\newcommand{\ea}{\end{align*}}

\renewcommand{\figurename}{Figure}

\newcommand{\norm}[1]{\left\lVert#1\right\rVert}

\begin{document}

\begin{frontmatter}

\title{Physics-Informed Neural Networks for Shell Structures}

\author{Jan-Hendrik Bastek}
\author{Dennis M. Kochmann\corref{mycorrespondingauthor}}
\address{Mechanics \& Materials Lab, Department of Mechanical and Process Engineering, ETH Zürich, 8092 Zürich, Switzerland}
\cortext[mycorrespondingauthor]{Corresponding author}
\ead{dmk@ethz.ch}

\begin{abstract}
The numerical modeling of thin shell structures is a challenge, which has been met by a variety of finite element (FE) and other formulations---many of which give rise to new challenges, from complex implementations to artificial locking. As a potential alternative, we use machine learning and present a Physics-Informed Neural Network (PINN) to predict the small-strain response of arbitrarily curved shells. To this end, the shell midsurface is described by a chart, from which the mechanical fields are derived in a curvilinear coordinate frame by adopting Naghdi's shell theory. Unlike in typical PINN applications, the corresponding strong or weak form must therefore be solved in a non-Euclidean domain. We investigate the performance of the proposed PINN in three distinct scenarios, including the well-known Scordelis-Lo roof setting widely used to test FE shell elements against locking. Results show that the PINN can accurately identify the solution field in all three benchmarks if the equations are presented in their weak form, while it may fail to do so when using the strong form. In the thin-thickness limit, where classical methods are susceptible to locking, training time notably increases as the differences in scaling of the membrane, shear, and bending energies lead to adverse numerical stiffness in the gradient flow dynamics. Nevertheless, the PINN can accurately match the ground truth and performs well in the Scordelis-Lo roof benchmark, highlighting its potential for a drastically simplified alternative to designing locking-free shell FE formulations.

\end{abstract}

\begin{keyword}
Structural mechanics\sep Machine learning\sep Shell theory
\sep Finite Element Method
\end{keyword}

\end{frontmatter}


\section{Introduction}

Shells are solids that are significantly thinner in one dimension compared to the other two. They serve as integral components of both natural and man-made structures. Technically speaking, shells are arbitrarily curved surfaces embedded in three-dimensional (3D) space, endowed with a (spatially variant or uniform) thickness normal to that surface. Therefore, shells can be efficiently approximated by a two-dimensional theory, capturing the out-of-plane kinematics by some simplifying assumptions, which are valid as long as the thickness is \textit{small} compared to the in-plane extensions. Due to their evident importance, shells have attracted considerable interest in the scientific community, especially since the second half of the last century, when Naghdi contributed what is often considered the standard reference for the theoretical treatment of shell structures \cite{Naghdi1973}. 

From a mathematical perspective, the mechanical problem is usually defined on the shell midsurface, a two-dimensional (2D) manifold embedded in the 3D physical space. The shell governing equations---in the form of a set of partial differential equations (PDEs) or the corresponding variational problem---must be solved on that manifold, which implies notable technical differences compared to the typically considered Euclidean space. Depending on the chosen shell model, the shape of the surface, and boundary conditions, analytical solutions can be found for a small class of special cases. In general and typical for engineering practice, however, one must rely on numerical methods to obtain solutions. To this end, the finite element method (FEM) is widely used in commercial \cite{Simulia2014,COMSOL2019} and open-source frameworks \cite{Hale2018}, though other approaches such as mesh-free methods exist as well \cite{Hale2012}. We note that the implementation of shell models into the FEM setting is technically complex, as they are deeply rooted in differential geometry. Furthermore, special care must be taken to ensure that the chosen finite element (FE) space is free from \textit{locking}, which arises if certain deformation modes cannot be represented. Addressing these limitations of mesh-based methods has led to a plethora of research over the past decades, as summarized in \cite{Chapelle2011}. Although the implementation of shell models has been considerably simplified in recent years \cite{Hale2018}, the choice of suitable shape functions and discretization techniques to prevent locking still requires significant user experience and/or model fine-tuning.

Recently, methods of machine learning (ML) and, more specifically, deep learning \cite{Lecun2015} have gained interest for applications in the natural sciences and engineering. Traditionally, such methods operate in a data-driven setting and may \textit{learn} system responses by identifying correlations based on a large dataset gathered from experiments or simulations. This can drastically accelerate subsequent evaluations to promote inexpensive surrogate models \cite{Zheng2021,Bastek2022}. In contrast to such methods stand the so-called \textit{Physics-Informed Neural Networks} (PINNs), which, although originally proposed decades ago \cite{Lagaris1998}, have only recently gained traction \cite{Raissi2019} due to advances in computing power. PINNs do not require any data to obtain the response of the physical system. Instead, the governing equations are directly incorporated into the loss function of the deep learning framework, so that minimization of the loss becomes equivalent to fulfilling the physical governing equations. PINNs have successfully been applied to diverse areas such as fluid mechanics \cite{Cai2022}, heat transfer \cite{Cai2021}, and solid mechanics \cite{Haghighat2021}. Alongside their empirical success, there is a growing body of theoretical underpinning, e.g., to bound errors \cite{Mishra2022} or to improve the rate of convergence \cite{Wang2022}.

When it comes to structural mechanics, prior work on PINNs has shown promising results for predicting the structural response of \textit{plates} \cite{Li2021,Zhuang2021} (which can be understood as the simplest case of initially \textit{flat} shells). Recently, this work was extended to composite plates and the special class of (shallow) cylindrical shells coupled to \textit{Extreme Learning Machines} with improved computational efficiency but reduced accuracy, especially if trained without external data \cite{Yan2022}.

Motivated by the reported success, we here present a PINN framework for simulating the mechanical response of arbitrarily curved continuous shells.
We evaluate the accuracy of the solutions obtained with this framework in comparison to FEM solutions. Although we foresee this framework to be of primary interest to the computational mechanics community, this study may also, in a more general sense, shed light on the performance of PINNs in non-Euclidean domains for a set of non-trivial physical equations and extends prior studies \cite{Tang2021}. As the basis for our formulation, we consider the \textit{linear Naghdi shell model}, suitable for the description of small-strain deformations of arbitrarily shaped shell structures. This model includes \textit{shear} deformation, which enables the accurate description of comparably \textit{thick} shells of significant relevance in engineering practice.

The remainder of this contribution is organized as follows. Naghdi shell theory, including a short primer on the necessary differential geometry concepts, is summarized in Section~\ref{section_naghdi}. In Section~\ref{section_PINN} we briefly introduce the general setting of PINNs and detail the application to Naghdi shell theory. We validate the framework by comparing it with reference FEM solutions in Section~\ref{section_results}, where we consider a representative set of three shell structures with fundamentally distinct curvatures and boundary conditions. Section~\ref{sec:Conclusions} concludes our study and discusses extensions and generalizations.

\section{Naghdi shell model}\label{section_naghdi}
\subsection{Geometrical preliminaries}\label{geom_measures}

We begin by introducing those concepts from differential geometry required to describe the shell kinematics. In general, a shell is a 3D body bounded by two outer surfaces and a \textit{thickness} as the distance between those two surfaces. Its shape can be described by the \textit{midsurface}, located in the middle of the two outer surfaces. Typically, the thickness of the shell is much smaller than the characteristic dimensions of the midsurface and thus admits a reduced kinematic parameterization in terms of the (2D) midsurface only. For better differentiation, in the following we reserve Latin indices for 3D quantities (i.e., $i,j,\ldots=1, 2, 3$), while Greek indices relate to the 2D midsurface (i.e., $\alpha,\beta,\ldots = 1,2$), based on which the majority of the necessary geometric measures can be derived. Superscripts $(\cdot)^i$ refer to contravariant components of a tensor, while subscripts $(\cdot)_i$ refer to covariant components. Furthermore, Einstein's summation convention is tacitly implied. Bold symbols indicate tensors of order one or higher. A single dot denotes a simple contraction, e.g., contracting two second-order tensors $\bfA, \bfB$ with base vectors $\bfa$ is written as
\be
    \bfA \cdot \bfB \defeq A^{\alpha \beta} B_{\beta \gamma} \bfa_{\alpha} \otimes \bfa^{\gamma},
\ee
while a colon denotes a double contraction, i.e.,
\be
    \bfA : \bfB \defeq A^{\alpha \beta} B_{\beta \alpha}.
\ee
In the following, we partly follow the derivation in \cite{Gaile2011} but focus only on the key steps necessary for our study; further details and explanations can be found in \cite{Ciarlet2005}.

\begin{figure}
	\centering
	\includegraphics[width=0.65\textwidth]{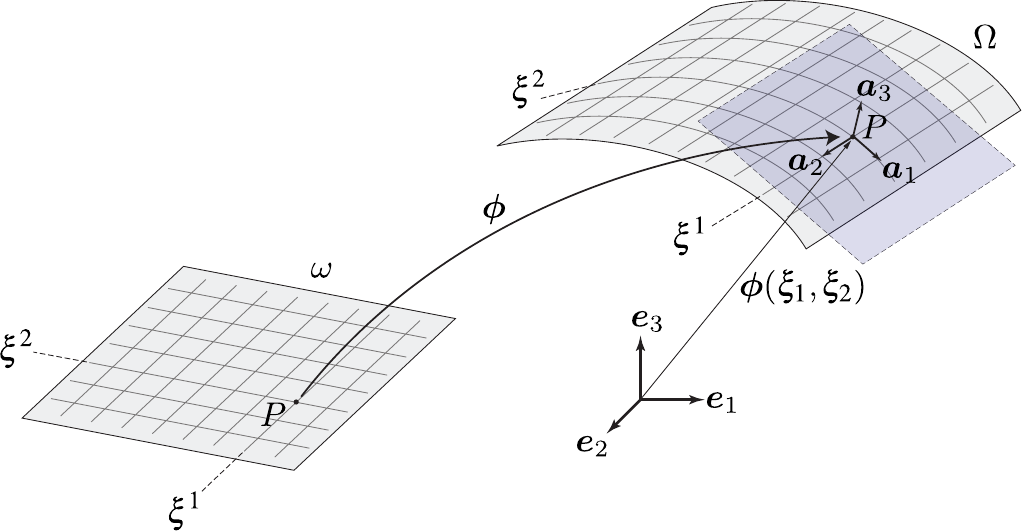}
	\caption{Definition of the shell midsurface based on the chart $\bm{\phi}$, which maps from the reference domain $\omega$ to the physical domain $\Omega$. Besides the global frame with basis $\{\bm{e}_1,\bfe_2,\bf_3\}$, we construct from $\bm{\phi}$ a local covariant basis $\{\bm a_1,\bfa_2,\bfa_3\}$ at any point $P$, with $\bfa_1$ and $\bfa_2$ spanning the local tangent plane and $\bfa_3$ being normal to the midsurface.}
	\label{fig:midsurface}
\end{figure}

Let us assume that the midsurface of a smooth continuous shell is given by a \emph{two-dimensional chart} $\bm{\phi}:\calR^2 \rightarrow \calR^3$, i.e., a smooth injective mapping from a reference domain $\omega \subset \calR^2$ into the physical space $\Omega \subset \calR^3$, as schematically shown in Figure~\ref{fig:midsurface}. We describe a position on the shell by curvilinear coordinates $(\xi^1,\xi^2)\in\omega$, which are defined on the manifold, so the midsurface is entirely defined by $\bm{\phi}(\xi^1,\xi^2)\in\Omega$.

To describe \textit{local} quantities, we construct a covariant basis at each point of the midsurface as
\be
\bm{a}_{\alpha} = \displaystyle \frac{\partial \bm{\phi}}{\partial \xi^{\alpha} } \quad\text{for }\alpha=1,2,\qquad
\bm{a}_{3} = \frac{\bm{a}_1 \times \bm{a}_2}{\norm{\bm{a}_1 \times \bm{a}_2}}.
\ee
From this basis, we retrieve the (covariant) components of the metric tensor,
\be
a_{\alpha \beta} =  \bm{a}_{\alpha} \cdot \bm{a}_{\beta},
\ee
which capture \textit{in-plane} deformation and rotations of a material point on the midsurface. 

When taking derivatives along a curve on the midsurface, we must account for the change of the local basis along that curve. For a vector field $\bm{v}=v_\alpha\bfa^\alpha$ defined on the midsurface, we thus obtain
\be\label{eq:deriv}
\frac{\partial \bm{v}}{\partial \xi^{\beta}} = v_{\alpha, \beta} \bm{a}^{\,\alpha}  +  v_{\alpha} \bm{a}^{\,\alpha}_{\,,\beta}
= v_{\alpha | \beta} \bm{a}^{\,\alpha} + b^{\alpha}_{\beta} v_{\alpha} \bm{a}_3,
\ee
where we introduced the (more standard) differential geometric notation with
the \textit{surface covariant derivative}
\be
v_{\alpha | \beta} =  v_{\alpha, \beta}  - \Gamma^{\rho}_{\alpha \beta} v_{\rho}
\ee
and the \textit{Christoffel symbol}
\be
\Gamma^{\rho}_{\alpha \beta} = \bm{a}_{\alpha, \beta} \cdot \bm{a}^{\,\rho}
\ee
as a measure for the in-plane curvature. The normal component is characterized by the \textit{second fundamental form}, which measures the \textit{extrinsic} (i.e., out-of-plane) curvature of the midsurface as
\be
b^{\alpha}_{\beta} = -\bm{a}_{3,  \beta} \cdot \bm{a}^{\, \alpha} = \bm{a}_3 \cdot \bm{a}^{\, \alpha}_{ ,\beta}.
\ee
Furthermore, we relate the convected (covariant) basis vectors $\{\bm{g}_1,\bfg_2,\bfg_3\}$, which are not necessarily parallel to the midsurface, to the midsurface basis vectors $\bm{a}_i$ via
\be\label{eq:basis_transformation}
\bm{g}_{\alpha} = \Big(\delta^{\lambda}_{\alpha} - \xi^3 b^{\lambda}_{\alpha}  \Big) \bm{a}_{\lambda},\quad\bm{g}_3 =  \bm{a}_3 \quad\text{so that} \quad \bfg_\alpha\cdot\bfg^\beta=\delta_\alpha^\beta
\ee
with the Kronecker delta $\delta_\alpha^\lambda$.
Lastly, we express the \textit{third fundamental form} as
\be
c_{\alpha  \beta} =  b^{\rho}_{\alpha} b_{\rho \beta}.
\ee
Note that in order to integrate over the shell in physical space, we may conveniently relate the physical surface area differential $\dd\mathcal{S}$ to the reference coordinates via the determinant of the metric tensor, $a=\det\bfa$ according to
\be
\dd\mathcal{S} = \norm{ \bm a_1 \times \bm a_2 } \dd\xi^1  \dd\xi^2 = \sqrt{\det\bm{a}} \dd\xi^1  \dd\xi^2 \defeq \sqrt{a} \dd\xi^1  \dd\xi^2.
\ee

\subsection{Shell model}\label{shell_model}

For the mechanical description of a shell structure, we adopt the conventions of the so-called \textit{5-parameter shell} model, for which the reader is referred to \cite{Naghdi1973} for a more comprehensive derivation.
We express the \textit{undeformed configuration} of a shell in 3D physical space as
\be
\bm{\Phi}(\xi^1, \xi^2, \xi^3) = \bm{\phi}(\xi^1, \xi^2) + \xi^3 \bm{a}_3(\xi^1, \xi^2), \quad (\xi^1, \xi^2) \in \omega \subset \mathbb{R}^2, \quad \xi^3 \in [-t/2,  t/2],
\ee
where we assume the thickness $t$ to be constant. To obtain the displacements, we assume that the normal fibers are inextensible and remain straight after deformation, which is also known as the \textit{Reissner-Mindlin kinematic assumptions}. As a consequence (and because the material fibers as assumed infinitesimally thin), we can express the rotation of each material fiber uniquely by a rotation vector normal to this fiber, i.e., via $\theta_{\lambda} \bm{a}^{\,\lambda}$. The overall displacement vector field is hence composed of a translation of points and the introduced rotation of the associated director as 
\be
\bm{U} = \bm{U}(\xi^1, \xi^2, \xi^3) = \bm{u}(\xi^1, \xi^2) + \xi^3 \theta_{\lambda}(\xi^1, \xi^2) \bm{a}^{\,\lambda}(\xi^1, \xi^2).
\ee
To measure deformations in the (small-strain) regime of linearized kinematics, we consider the \textit{linearized Green-Lagrange strain tensor}
\be\label{eq:linear_green_lagrange}
E_{ij} = \frac{1}{2} \big(\bm{g}_i \bm{U}_{,j} + \bm{g}_j \bm{U}_{,i} \big),
\ee
whose components evaluate to (see \ref{app:Derivation})
\begin{align} 
\begin{split}
E_{\alpha \beta} &= \frac{1}{2} \left( u_{\alpha | \beta} + u_{\beta | \alpha} \right) - b_{\alpha \beta} u_3 
 +\xi^3 \left[\frac{1}{2} \bigg( \theta_{\alpha | \beta} + \theta_{\beta | \alpha} - b^{\lambda}_{\beta} u_{\lambda | \alpha } - b^{\lambda}_{\alpha} u_{\lambda | \beta } \bigg) + c_{\alpha \beta} u_3\right]\\
 & \quad + \frac{(\xi^3)^2}{2}  \left( b^{\lambda}_{\beta} \theta_{\lambda | \alpha } + b^{\lambda}_{\alpha} \theta_{\lambda | \beta } \right),
\end{split}\\
E_{\alpha 3} &= \frac{1}{2}(\theta_{\alpha} + u_{3, \alpha} + b^{\lambda}_{\alpha} u_{\lambda}),\\
E_{33} &= 0.
\end{align}
As the shell is assumed to be \textit{thin}, those terms that are quadratic in $\xi^3$ are neglected, and we retrieve the strain contributions of the Naghdi shell model as
\be\label{eq:strain_measures}
\begin{split}
    &e_{\alpha\beta}(\bm u) = \displaystyle \frac{1}{2} \big( u_{\alpha | \beta} + u_{\beta | \alpha} \big)
    - b_{\alpha \beta} u_3,\\
    &k_{\alpha\beta}(\bm u, \bm \theta) = \displaystyle \frac{1}{2} \big( \theta_{\alpha | \beta} + 
    \theta_{\alpha | \beta} - b^{\lambda}_{\beta} u_{\lambda | \alpha} -  b^{\lambda}_{\alpha} u_{\lambda | \beta} \big) + c_{\alpha \beta} u_3,\\
    &\gamma_{\alpha}(\bm u, \bm \theta) = \displaystyle \theta_{\alpha} + u_{3, \alpha} + 
    b^{\lambda}_{\alpha} u_{\lambda},
\end{split}
\ee
where $e_{\alpha \beta}(\bm u)$, $k_{\alpha\beta}(\bm u, \bm \theta)$, and $\gamma_{\alpha}(\bm u, \bm \theta)$ are interpreted as \textit{membrane}, \textit{bending}, and \textit{shear} strains, respectively. Owing to the variational structure of the problem, the total potential energy functional of a homogeneous, isotropic, linear elastic shell becomes
\be\label{eq:energy}
    \Pi[\bm u, \bm \theta] = \displaystyle \frac{1}{2} \int_{\omega} \left[\mathbb{C}^{\alpha \beta \sigma \rho} \left( t e_{\alpha \beta} e_{\sigma \rho} + \frac{t^3}{12} k_{\alpha \beta} k_{\sigma \rho} \right) + \mathbb{D}^{\alpha \beta} t \kappa \gamma_{\alpha} \gamma_{\beta} \right]\dd\mathcal{S} - W_{\text{ext}},
\ee
with fourth- and second-order tensor components
\begin{align}\label{eq:elasticity_tensors}
    &\mathbb{C}^{\alpha \beta \sigma \rho} = \displaystyle \frac{2 \lambda \mu}{\lambda + 2\mu} a^{\alpha \beta} a^{\sigma \rho} + \mu(  a^{\alpha \sigma }a^{\beta \rho} + a^{\alpha \rho  }a^{\beta \sigma}),\\[4pt]
    \label{shear_law} &\mathbb{D}^{\alpha \beta} = \mu a^{\alpha \beta},
\end{align}
where $\lambda$ and $\mu$ are the Lam\'{e} elastic moduli. $W_\text{ext}$ represents the work done by external forces.
For further details we refer to \cite{Naghdi1973,Gaile2011} and only emphasize that the elasticity tensors in \eqref{eq:elasticity_tensors} are derived under plane-stress conditions through the thickness, while the shear correction factor $\kappa$ accounts for the nonlinear distribution of shear strains across the thickness. For a homogeneous, isotropic base material, its value is taken as $\kappa = 5/6$. 

To identify the equilibrium configuration for a given set of essential boundary conditions and external loading, we may leverage the principle of minimum potential energy,
\be\label{eq:min_energy}
    \{\bm{u}^*, \bm{\theta}^*\} = \argmin \; \left\{\Pi[\bm{u}, \bm{\theta}] \ \text{s.t. essential BCs}\right\},
\ee
which constitutes the basis for the weak form used in the FE setting. Of course, we can equivalently express the equilibrium conditions in terms of the \textit{strong form}, as presented in detail in \cite{Chapelle1998}. To this end, we denote the membrane and shear force tensor as well as the bending moment tensor by, respectively,
\be
\begin{split}
\bm n &= t \, \mathbb{C}:\bm e, \qquad
\bm q = t \, \mathbb{D} \cdot \bm \gamma, \qquad
\bm m = \frac{t^3}{12} \mathbb{C}:\bm k.
\end{split}
\ee
The equilibrium equations in $\omega$ follow as
\be\label{eq:shell_strong_form}
\begin{split}
    \divv{\bm m} - \bm  q &= \bm 0,\\
    \divv(\bm n- \bm b\cdot \bm m) - \bm b\cdot \bm q + \bm f &= \bm 0,\\
    \divv \bm  q + \bm b : (\bm n - \bm b\cdot \bm m) + f_3 &= 0,
\end{split}
\ee
with the natural boundary conditions
\be\label{eq:shell_natural_bc}
\begin{split}
    \bm m \cdot \bm\nu  &= \bm 0,\qquad
    (\bm n-\bm b\cdot \bm m) \cdot \bm\nu = \bm 0,\qquad
    \bm q \cdot \bm\nu = 0
\end{split}
\ee
on the free boundary $\partial\omega_{\text{N}}$, where $\bm\nu$ is the unit outward normal perpendicular to the tangent plane. $\bff$ represents the distributed body force tangential to the shell surface, and $f_3$ the out-of-plane force component.
Note that the divergence must again consider the curvilinear base and is obtained by contracting the covariant surface derivative on the last two indices, i.e., for a vector $\bm a$ and second-order tensor $\bm B$ we have
\be
\begin{split}
    \divv \bm a &= a^{\alpha}_{|\alpha},\qquad
    \divv \bm B = B^{\alpha\beta}_{|\beta}.
\end{split}
\ee

Note that we here consider the so-called \textit{5-parameter model}, which has five degrees of freedom: the three displacement field components $\bm u\in\mathbb{R}^3$ and the two rotation components $\bm\theta\in\mathbb{R}^2$, since rotations of the material fibers around their own axis (also referred to as \textit{drilling}) are neglected. To prevent this by construction, we directly solve for the rotations with respect to the two base vectors of our local basis, which are tangential to the surface (i.e., $\bm a_1$ and $\bm a_2$). The displacement field is instead defined in the global Cartesian frame denoted by $\bm{\hat{u}}\in \mathbb{R}^3$ and must thus first be transformed to the local frame via the linear mapping
\be
    \bm{u} = \bm{T} \bm{\hat{u}}, \quad
    \bm{T} = \begin{pmatrix}
    | & | & |\\
    \bm{a}_1 & \bm{a}_2 & \bm{a}_3\\
    | & | & |
    \end{pmatrix}\T,
\ee
before assembling the corresponding strains in \eqref{eq:strain_measures}.
Lastly, note that $\bm{T}$ necessarily depends on position $\bm\xi$, so care must be taken when evaluating the covariant derivatives of $\bm u$ via $\bm{\hat{u}}$.

\section{Physics-informed neural network framework} \label{section_PINN}

In this section, we present the details of the PINN framework. We first give a brief introduction to the general PINN methodology, before providing details of the application in the above shell setting.

\subsection{General PINN setting}
Consider the general form of a system governed by PDEs on a domain $\Omega \subset \mathbb{R}^{d}$, abbreviated as
\be\label{eq:PDE_set}
    \mathcal{F}[\bm{y}(\bm{x})]=\bm{0}, \quad \bm{x}=\left(x_{1}, x_{2}, \cdots, x_{d}\right) \in \Omega, \quad \bm{y}=\left(y_{1}(\bm{x}), y_{2}(\bm{x}), \cdots, y_{n}(\bm{x})\right) \in \mathbb{R}^n,
\ee
and boundary conditions
\be\label{eq:boundary_set}
    \mathcal{B}[\bm{y}(\bm{x})]= \bm{0}, \quad \bm{x} \in \partial \Omega,
\ee
where $\mathcal{F}$ is a differential operator, $\mathcal{B}$ is a boundary condition operator, $\partial\Omega$ is the boundary of the domain, and $\bm y$ is the solution of the set of PDEs for all $\bfx\in\Omega$. The fundamental idea of PINNs is to approximate $\bm y$ via a feed-forward multi-layer neural network $\mathcal N$, which is parameterized by weights and biases collected in $\bm\tau$, i.e.,
\be
    \bm y(\bm x) \approx \mathcal N(\bm x;\bm\tau) = \mathcal{N}_{\bm\tau}(\bm x).
\ee
To identify the set of hyperparameters $\bm\tau$ that best fulfills \eqref{eq:PDE_set} and \eqref{eq:boundary_set}, the PINN is trained to minimize the residual arising from not exactly fulfilling these equations. Notably, we can employ automatic differentiation (typically used to optimize the internal weights and biases of the network) to obtain derivatives appearing in \eqref{eq:PDE_set} and \eqref{eq:boundary_set} by evaluating the partial derivative of the output with respect to the input coordinates \cite{Raissi2019}, provided that the chosen activation function of our PINN is sufficiently smooth. To numerically estimate the residual, each equation is evaluated at a total of $N_{\text{c}}$ sampled \textit{collocation points}, the average of which is taken as the loss function of the network to be minimized by first- or second-order optimizers such as Adam \cite{Kingma2015} or L-BFGS \cite{Liu1989}, respectively. For further details, the reader is referred to \cite{Raissi2019}.

Note that certain classes of PDEs (including the present shell model, as explained above) possess a \textit{variational structure}, i.e., the solution $\bm y$ can also be interpreted as an extremal point of a functional $I[\bm y]$. Instead of directly enforcing the strong form given by \eqref{eq:PDE_set} and \eqref{eq:boundary_set}, we can thus equivalently minimize $I[\bm y]$, which has sometimes been referred to as the \textit{Deep Ritz Method} \cite{E2018}. This relaxes the conditions on the solution space, as it reduces the order of the highest derivatives appearing in the governing equations.

\subsection{Application to shells}

\begin{figure}
	\centering
	\includegraphics[width=1.0\textwidth]{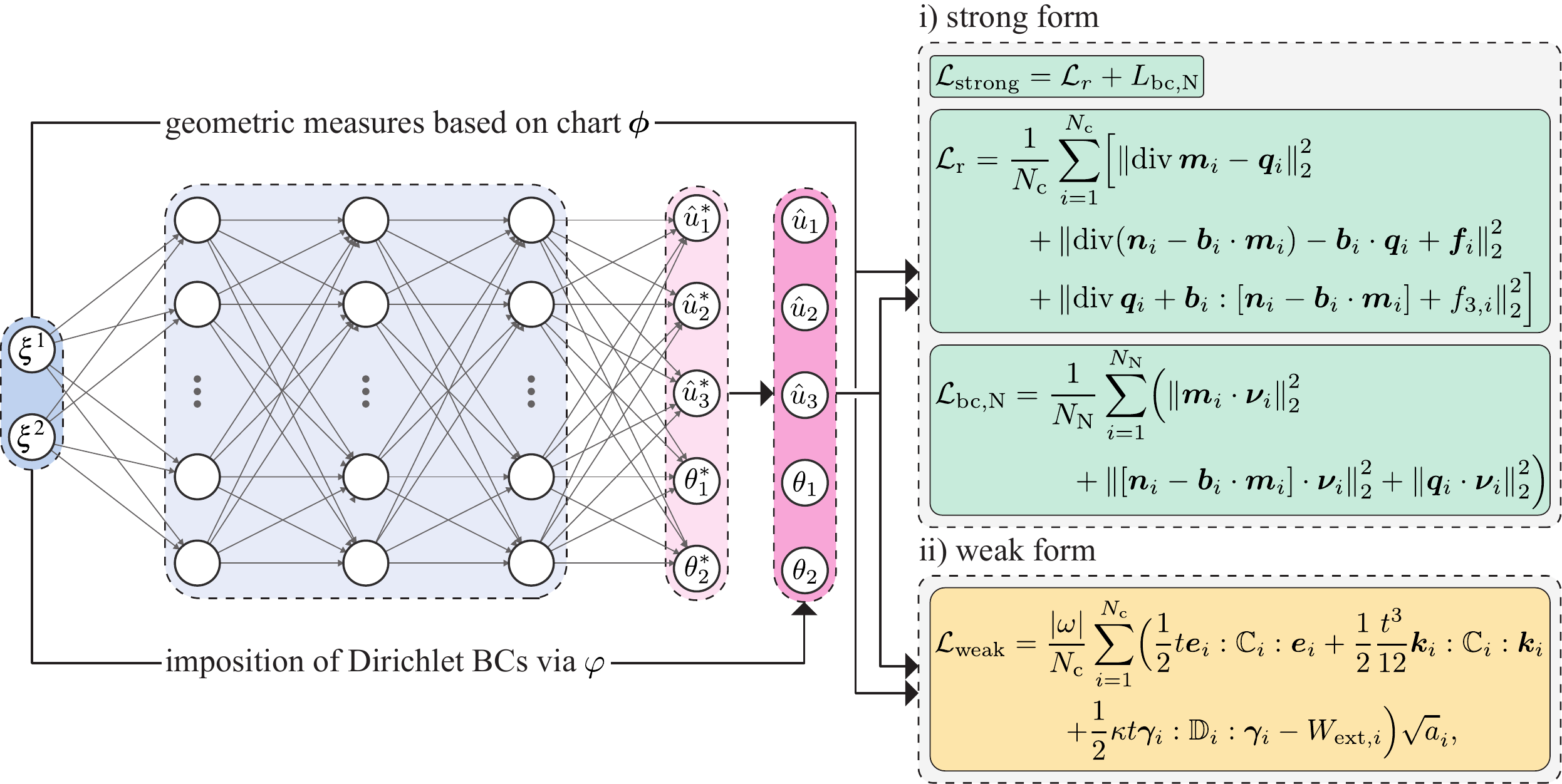}
	\caption{\textbf{PINN architecture and corresponding loss.} The PINN predicts the (scaled) global deformations $\hat{\bfu}^*$ and rotations $\bm{\theta}^*$ at a collocation point $\bm\xi_i$, which are multiplied by the trial function $\varphi$ to impose the Dirichlet BCs. Based on these five parameters and the given chart $\phi$, the shell equations can be assembled equivalently in their strong or weak forms, which define the loss function for the training of the network.}
	\label{fig:PINN_architecture}
\end{figure}

As derived in Section~\ref{shell_model}, the governing system of PDEs 
in the case of shell structures is given by \eqref{eq:shell_strong_form} and \eqref{eq:shell_natural_bc}. For a given set of boundary conditions, our PINN takes as input a set of curvilinear coordinates $\bm\xi\in \omega \subset \mathbb{R}^2$ and predicts the corresponding (global) displacements $\bm{\hat{u}}\in\mathbb{R}^3$ and (local) rotations $\bm\theta\in\mathbb{R}^2$ over the given (reference) domain $\omega$, from which the strains, forces/moments and, subsequently, the strong form can be constructed. Note that for \textit{natural} boundary conditions (relating to applied forces or moments) defined on $\partial \omega_{\text{N}}\subseteq\partial\omega$ the solution must satisfy \eqref{eq:shell_natural_bc}, while for \textit{essential} boundary conditions defined on $\partial \omega_{\text{D}}\subseteq\partial\omega$ we impose
\be
\begin{split}
    \bm{\hat{u}}(\bm\xi) &= \bm{\bar{u}}(\bm\xi) 
    \quad\text{and}\quad
    \bm\theta(\bm\xi) = \bm{\bar{\theta}}(\bm\xi), \quad\text{for } \bm\xi \in \partial \omega_{\text{D}},
\end{split}
\ee
where $\bm{\bar{u}}(\bm\xi)$ and $\bm{\bar{\theta}}(\bm\xi)$ are the prescribed displacements and rotations, respectively, and $\partial \omega=\partial \omega_{\text{D}} \cup \partial \omega_{\text{N}}$, and $\partial \omega_{\text{D}} \cap \partial \omega_{\text{N}} = \emptyset$ with $\partial\omega$ denoting the outer boundary of $\omega$.
When considering the strong form \eqref{eq:shell_strong_form} and \eqref{eq:shell_natural_bc}, the residual or, equivalently, the loss $L_{\text {strong}}$ of the PINN is estimated by using a measure of the deviation such as the \textit{mean-squared error}, which gives the total loss based on the strong form as
\be\label{eq:total_loss_strong}
    \calL_{\text{strong}}(\tau) = \calL_{\text{r}}(\tau)+ \calL_{\text{bc}}(\tau) = \calL_{\text{r}}(\tau)+\calL_{\text{bc,N}}(\tau) + \calL_{\text{bc,D}}(\tau),
\ee
where, with $\bm\xi_i=(\xi^1,\xi^2)_i$ denoting the location of the $i$th collocation point,
\be
    \begin{aligned}
    \calL_{\text{r}}(\tau) =\frac{1}{N_{\text{c}}} \sum_{i=1}^{N_{\text{c}}}& \left[\norm{\divv{\bm m_{\tau}(\bm\xi_i)} - \bm  q_{\tau}(\bm\xi_i)\large}_{2}^{2}
    + \norm{\divv(\bm n_{\tau}(\bm\xi_i)-\bm b(\bm\xi_i)\cdot \bm m_{\tau}(\bm\xi_i)) - \bm b(\bm\xi_i)\cdot \bm q_{\tau}(\bm\xi_i) + \bm f(\bm\xi_i)}_{2}^{2}\right.\\
    & \quad +  \left.\norm{\divv \bm  q_{\tau}(\bm\xi_i) + \bm b(\bm\xi_i) : [\bm n_{\tau}(\bm\xi_i) - \bm b(\bm\xi_i)\cdot \bm m_{\tau}(\bm\xi_i)] + f_3(\bm\xi_i)}_{2}^{2}\right]\\
    \end{aligned}
\ee
and
\begin{align}
    \calL_{\text{bc,N}}(\tau) &=\frac{1}{N_{\text{N}}} \sum_{i=1}^{N_{\text{N}}}\Bigl[\norm{\bm m_{\tau}(\bm\xi_i) \cdot \bm\nu(\bm\xi_i)}_{2}^{2} + \norm{[\bm n_{\tau}(\bm\xi_i)-\bm b(\bm\xi_i)\cdot \bm m_{\tau}(\bm\xi_i)] \cdot \bm\nu}_{2}^{2} + \norm{\bm q_{\tau}(\bm\xi_i) \cdot \bm\nu(\bm\xi_i)}_{2}^{2}\Bigr], \label{eq:vonNeumannBC}\\
    \calL_{\text{bc,D}}(\tau) &=\frac{1}{N_{\text{D}}} \sum_{i=1}^{N_{\text{D}}}\Bigl[\norm{\hat{\bm u}_{\tau}(\bm\xi_i)-\bm{\bar{u}}(\bm\xi_i)}_{2}^{2} + \norm{\bm \theta_{\tau}(\bm\xi_i)-\bm{\bar{\theta}}(\bm\xi_i)}_{2}^{2}\Bigr]. \label{eq:DirichletBC}
\end{align}
Note that sufficiently large numbers of collocation points for the residuals in the domain, on the natural and essential boundaries, denoted by $N_{\text{c}}$, $N_{\text{N}}$, and $N_{\text{D}}$, respectively, must be sampled to offer a good approximation of the total residual. The selection of the collocation points and weights can be random or according to some low-discrepancy sequence such as the Sobol sequence for a faster rate of convergence \cite{Sobol1967}, which was chosen here unless explicitly stated otherwise.

As the loss function \eqref{eq:total_loss_strong} consists of a total of 15  terms (five equilibrium equations, five natural boundary conditions, and five essential boundary conditions), it can be further simplified by directly imposing the essential boundary conditions on the PINN output. For example, consider the simple case of a square plate that is clamped on one edge, e.g.,
\be
\bm{\bar{u}}(\bm\xi)=\bm{0} \ \wedge \ \bm{\bar{\theta}}(\bm\xi)=\bm{0} \; \; \forall \ \xi^1=0, \ \xi^2 \in [0,1].
\ee
Defining the \textit{trial function} $\varphi(\bm\xi)=\xi^1$ to be multiplied with the PINN output will trivially fulfill this essential boundary condition and thus allows us to omit \eqref{eq:DirichletBC} in this special case. For more complex geometries, general frameworks to construct suitable distance functions exist and can be extended to also satisfy the natural boundary condition by construction \cite{Sukumar2022}.

As indicated before, we may equivalently leverage the variational structure of the problem and, instead of solving the strong form, minimize the potential energy functional \eqref{eq:min_energy}. In this case, an approximation of the total potential energy and hence the loss is given by
\be
\begin{split}
    \calL_{\text {weak}}(\tau) =\frac{|\omega|}{N_{\text{c}}} &\sum_{i=1}^{N_{\text{c}}}\Bigl(\displaystyle\underbrace{\frac{1}{2} t \bfe_{\tau}(\bm\xi_i):\mathbb{C}(\bm\xi_i):\bfe_{\tau}(\bm\xi_i)}_{\textrm{membrane energy}} + \underbrace{ \displaystyle \frac{1}{2}\frac{t^3}{12}\bfk_{\tau}(\bm\xi_i) : \mathbb{C}(\bm\xi_i) : \bfk_{\tau}(\bm\xi_i)}_{\textrm{bending energy}}\\
    \qquad+\displaystyle &\underbrace{\frac{1}{2}\kappa t \bm\gamma_{\tau}(\bm\xi_i) \cdot \mathbb{D}(\bm\xi_i) \cdot \bm\gamma_{\tau}(\bm\xi_i)}_{\textrm{shear energy}} - \displaystyle\underbrace{W_{\text{ext},\tau}(\bm\xi_i)}_{\textrm{external work}}\Bigr)\sqrt{a(\bm\xi_i)},
\end{split}
\ee
where $|\omega|=\int_\omega \dd\xi^1\dd\xi^2$ denotes the total area of the reference domain. This formulation offers various advantages: besides the reduced highest order of differentiation (saving computational costs), natural boundary conditions must not be incorporated by additional equations and thus further simplify the loss function to a single term (without potential ambiguity about the weighting of the various terms in the loss function).

Lastly, note that we can leverage automatic differentiation to not only assemble the loss function and train the network but also to efficiently and accurately evaluate all geometric quantities such as the fundamental forms introduced in Section~\ref{geom_measures} directly at the corresponding collocation points, which entirely depend on the prescribed chart $\bm{\phi}(\bm\xi)$. Since these quantities remain constant in the considered small-strain setting, we evaluate and store those before the training stage of our PINN and subsequently save computational resources. This is in contrast to mesh-based methods, in which these quantities are typically evaluated at quadrature points via interpolation between the element nodes, which, depending on the given chart and interpolation, may lead to a loss in accuracy for insufficiently refined meshes in subdomains with high local curvature. Besides, the computation of the out-of-plane normal may require additional averaging schemes between adjacent elements \cite{Gaile2011}. 

\subsection{Implementation details}

As with any deep learning framework, there is a large variety of hyperparameters to consider, ranging from the choice of network architecture (i.e., the number of layers and neurons per layer as well as the activation function) to the training parameters (i.e., the number of collocation points that the strong or weak form is evaluated on, weight initialization, artificial weights on different loss terms) to the choice of the optimizer. In the context of PINNs, numerous studies on the influence of such hyperparameters can be found in the literature \cite{Raissi2019,Zhuang2021,Jagtap2020}. We summarize that a fine-tuned network architecture, evaluated on an increasing number of collocation points, improves the accuracy of the solution to some extent. Nevertheless, such improvements are typically rather marginal for a reasonable choice of hyperparameters and may not be able to mitigate some of the more fundamental failures of PINNs previously observed in certain scenarios, e.g., when solving the Allen-Cahn equation with sharp transitions in the solution fields. In such cases, more elaborate strategies such as adaptive sampling must be pursued to improve the accuracy of the predicted solutions to an acceptable level \cite{Wight2020,Krishnapriyan2021}.
Our investigations confirm the general trend observed in the literature with regards to the choice of hyperparameters, so that we will omit in-depth parameter studies for which we refer the reader to, e.g., \cite{Zhuang2021}, and only highlight interesting observations where deemed appropriate. For all subsequent studies, we use a standard feedforward multi-layer perceptron with three hidden layers and 50 neurons per layer using the GELU activation function \cite{Hendrycks2016}, which was empirically found to give the best results. We used Pytorch \cite{Paszke2019} for the implementation and its default version of the L-BFGS solver \cite{Liu1989} with enabled Strong-Wolfe conditions to optimize the networks.

\begin{figure}
	\centering
	\includegraphics[width=1.0\textwidth]{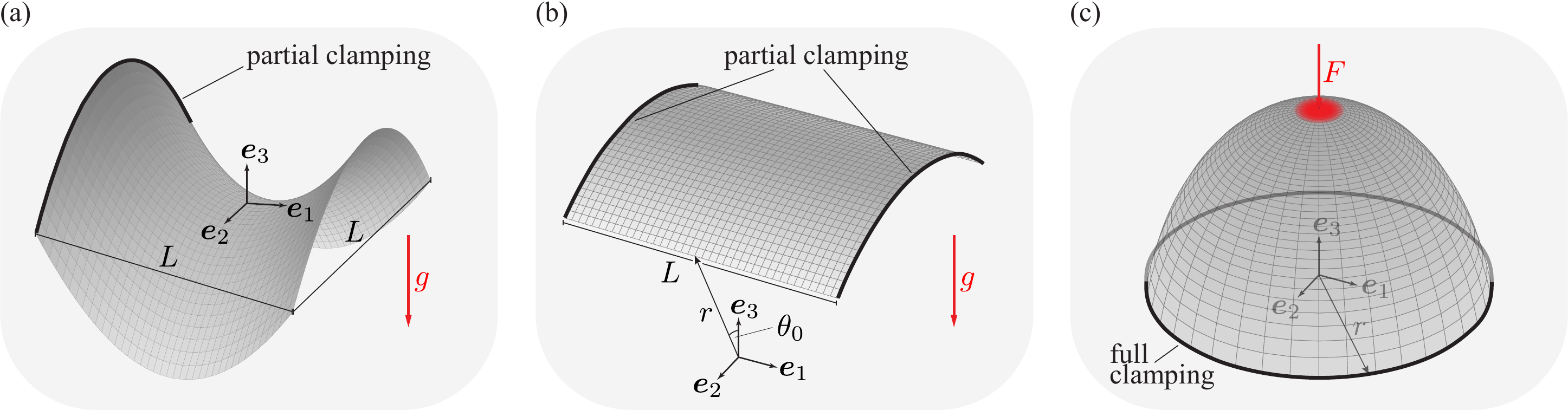}
	\caption{\textbf{Considered case studies.} a) A hyperbolic paraboloid fully clamped on one edge and subject to gravity loading. b) The Scordelis-Lo Roof benchmark \cite{Belytschko1985}, consisting of a partly clamped cylindrical shape subject to gravity loading ($\theta_0=40^{\circ}$). c) A fully clamped hemisphere subject to a vertical load at its center, modeled by a Gaussian kernel.}
	\label{fig:case_studies}
\end{figure}

\section{Results}\label{section_results}

Having derived the general setup of our PINN applied to shell theory, we proceed to test its performance on three fundamentally distinct shell problems, addressing hyperbolic, parabolic, and elliptic shapes as well as different boundary conditions, as shown in \figurename~\ref{fig:case_studies}. Specifically, we consider (a)~a hyperbolic paraboloid clamped at one edge and subject to gravity loading, (b)~a cylinder segment clamped at two edges and subject to gravity loading (this is also known as the Scordelis-Lo roof benchmark in the context of FE frameworks \cite{Belytschko1985}), and (c)~a fully clamped hemisphere subject to a concentrated Gaussian load. Since all benchmarks are in the realm of linear elasticity and geometrically linear theory, the choice of Young's modulus $E$ and magnitude $f$ of the applied load simply rescales the solution fields. Therefore, without loss of generality, we set $E=f=1$ for all subsequent studies to not artificially inflate the magnitudes of the different terms at the beginning of the training, which has been shown to deteriorate the convergence of PINNs \cite{Wang2021}.

To validate the results, we perform FE simulations of all three benchmarks, using the open-source library FEniCS-Shells \cite{Hale2018}, which offers a comparably simple implementation of the typically rather complex shell formulations. For all presented results, the FE mesh was refined until convergence was observed.

\subsection{Partially-clamped hyperbolic paraboloid}

We first consider a shell with the midsurface defined by the chart $\bm{\phi}(\xi^1,\xi^2)=\{ \xi^1,\xi^2,(\xi^1)^2-(\xi^2)^2 \}$ with $(\xi^1,\xi^2) \in [-1/2,1/2]$ (i.e., $L=1$), also referred to as a \textit{hyperbolic paraboloid} \cite{Hale2018}, which has a negative mean Gaussian curvature. As stated above, we can efficiently pre-compute all necessary geometric measures such as the fundamental forms based on the chart (and its derivatives with respect to $\xi^1,\xi^2$) by leveraging Pytorch's autograd engine, which gives us the \textit{exact} quantity at the chosen collocation points via automatic differentiation.

The one-sided clamping is formally expressed by $\hat{\bm{u}}=\bm{0}$ and $\bm{\theta}=\bm{0}$ for all $\xi^1=-1/2$, so that we define the trivial trial function $\varphi(\bm\xi)=\xi^1+1/2$ to fulfill these Dirichlet boundary conditions. The gravity load is implemented in the strong form and enters the weak form as $W_{\text{ext}}=-\int_{\omega}\rho g t\hat{u}_3 \dd \calS$ with $\rho g=1$ (as explained above). We set Poisson's ratio to $\nu=0.3$ and consider a characteristic thickness of $t/L=0.1$.

For both strong and weak forms, $2{,}048$ collocation points were sampled in the reference domain. Additionally, the strong form requires the explicit evaluation of the natural boundary conditions on the three unclamped edges, for which an additional $512$ collocation points each were sampled.

\begin{figure}
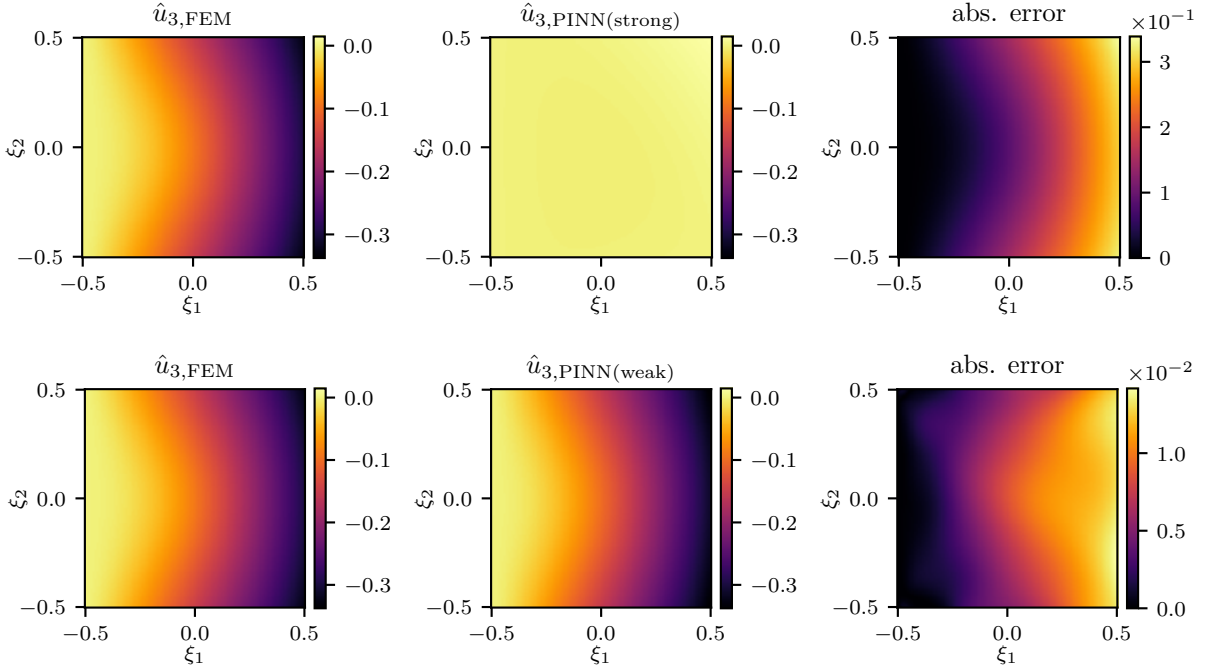

    \begin{center}
        \scalebox{1.0}{\import{figures/hyperb_parab/comparison_sol}{fig0.pgf}}
        \scalebox{1.0}{\import{figures/hyperb_parab/comparison_sol}{fig1.pgf}}
    \end{center}
    \caption{Comparison of the predicted $\hat{u}_3$-displacement field as obtained from the FE and PINN frameworks (based on both the strong and weak forms with $N_{\text{c}}=2{,}048$) for the partly-clamped hyperbolic paraboloid subject to gravity loading, shown in the reference domain. Displacements are scaled by a factor of $0.005$ for consistency with Figure~\ref{fig:3d_sol_hyperb_parab}.}\label{fig:u3_hyperb_parab}
\end{figure}

\begin{figure}
\centering
\begin{subfigure}[b]{0.4\textwidth}
\centering
\begin{tikzpicture}[scale=0.75]
\begin{axis}[
    label style={font=\large},
    xlabel={Epoch},
    ylabel={Average rel.\ $L_2$-error},
    xmin=0, xmax=100,
    ymin=0.01, ymax=4.0,
    ymode=log,
    legend pos=south west,
    legend cell align=left,
    ymajorgrids=true,
    grid style=dashed,
]
\addplot  [color=blue]
    table [x index=0, y index=1, col sep=comma] {figures/plots/L2_error_strong.csv};

\addplot  [color=red]
    table [x index=0, y index=1, col sep=comma] {figures/plots/L2_error_weak.csv};

\addplot  [color=green]
    table [x index=0, y index=1, col sep=comma] {figures/plots/L2_error_weak_fine.csv}; 

\legend{Strong form, Weak form, Weak form$^*$}
  
\end{axis}
\end{tikzpicture}
\caption{\normalsize{\textit{Partly} clamped hyperbolic paraboloid}}\label{se1}
\end{subfigure}
\centering
\begin{subfigure}[b]{0.49\textwidth}
\begin{tikzpicture}[scale=0.75]
\begin{axis}[
    label style={font=\large},
    xlabel={Epoch},
    ylabel={Average rel.\ $L_2$-error},
    xmin=0, xmax=100,
    ymin=0.01, ymax=4.0,
    ymode=log,
    legend pos=north east,
    legend cell align=left,
    ymajorgrids=true,
    grid style=dashed,
]

\addplot  [color=blue]
    table [x index=0, y index=1, col sep=comma] {figures/plots/L2_error_strong_fc.csv};

\addplot  [color=red]
    table [x index=0, y index=1, col sep=comma] {figures/plots/L2_error_weak_fc.csv};

\legend{Strong form, Weak form}
\end{axis}
\end{tikzpicture}
\caption{\normalsize{\textit{Fully} clamped hyperbolic paraboloid}}\label{daasd}
\end{subfigure}
\caption{\label{fig:loss_convergence} Average relative $L_2$-error of the five solution fields, computed as $\frac{1}{5}\left(\sum_{i=1}^3\sqrt{ (u_{i,\text{FEM}}-u_{i,\text{PINN}})^2/u_{i,\text{FEM}}^2}+\sum_{j=1}^2\sqrt{ (\theta_{j,\text{FEM}}-\theta_{j,\text{PINN}})^2/\theta_{j,\text{FEM}}^2}\right)$, with respect to the FEM solution over 100 training epochs based on the strong and weak forms. We first consider the original problem, i.e., a \textit{partly} clamped shell, and in addition investigate the performance of both the strong and weak forms for the simpler problem of a \textit{fully} clamped shell. $^*$Trained using $N_{\text{c}}=16{,}384$ (other results for $N_{\text{c}}=2{,}048$).}
\end{figure}
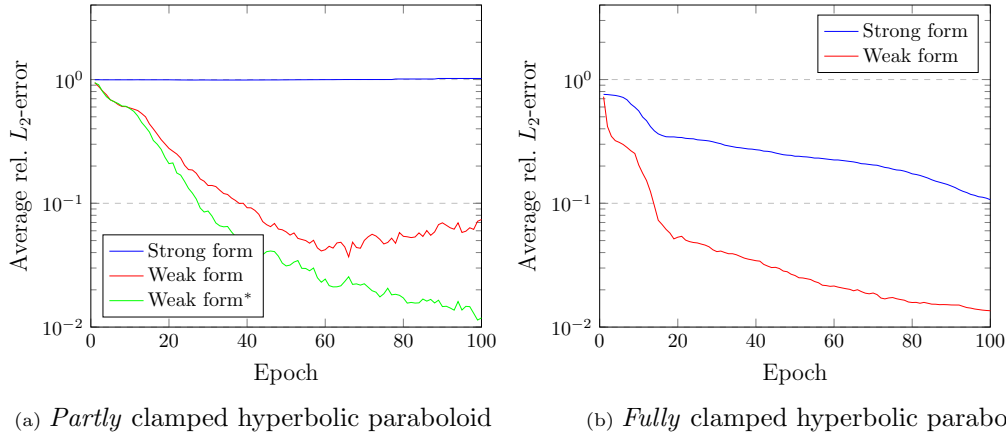

We compare the results based on the PINNs applied to both the strong and the weak forms with the results from FEniCS in \figurename~\ref{fig:u3_hyperb_parab}, where the solution for the $\hat{u}_3$-displacement field is evaluated over the reference domain. For analogous comparisons of the other four solution fields (based on the weak form only, since the strong form offers no further insights) we refer to \ref{appdx_hyperb}. Both the strong and weak form were trained for 100 epochs, at which the loss has approximately converged, as shown in \figurename~\ref{fig:loss_convergence}a.

While the PINN using the strong form as the loss function fails to converge to the FEM solution, the PINN based on the weak form succeeds at accurately matching the solution of FEniCS. This confirms the trend observed in \cite{Li2021}, there for the simpler case of a plate. Interestingly, we observed an \textit{increase} in the average $L_2$-error for the weak form using $N_{\text{c}}=2{,}048$ collocation points after around 60 epochs, as the PINN identified a configuration with smaller total potential energy than the physical solution due to a too coarse sampling of the collocation points. Increasing the number of collocation points to larger values (we selected, e.g., $N_{\text{c}}=16{,}384$) does consistently remove this effect. We generally find that the inherent over-parameterization of the network may lead to such non-physical artifacts, if the sampled collocation points are too coarse. To furthermore explain why the weak form greatly outperforms the strong form, we refer to its two obvious advantages: besides reducing the required order of differentiation, natural boundary conditions are directly incorporated and must not be expressed by additional terms in the loss function, which facilitates the training.

\begin{figure}
    \begin{center}
        \import{figures/hyperb_parab/3d_sol_fc}{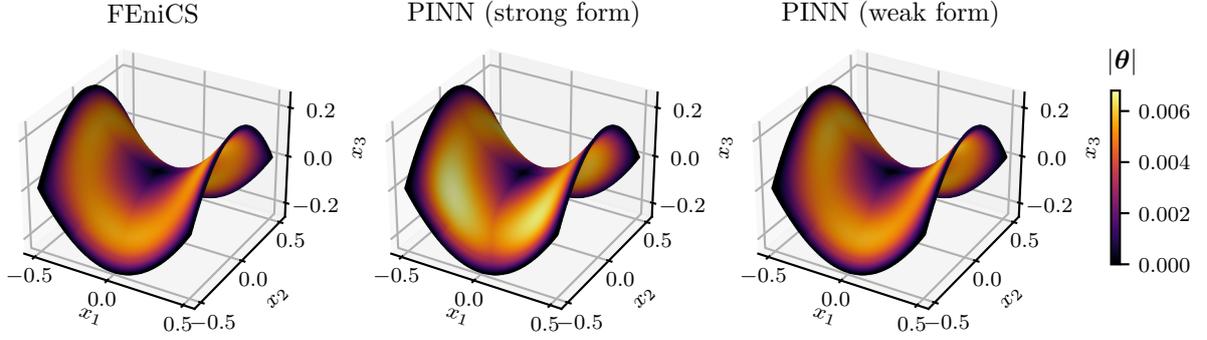}
    \end{center}
    \vspace{-6mm}
    \caption{Comparison of the deformation predicted by the FE and PINN framework (based on both the strong and weak forms with $N_{\text{c}}=2{,}048$) for the \textit{fully} clamped hyperbolic paraboloid subject to gravity load in the physical space, shown in its deformed configuration. The surface color corresponds to the norm of the rotation fields, $|\bm{\theta}|=|\theta_1|+|\theta_2|$. Displacements are scaled by a factor of $0.005$ for improved visibility.}\label{fig:3d_sol_hyperbolic_parab_fc}
\end{figure}

We hypothesize that it is particularly the extended loss function (with five additional terms) that contributes to the failure of the strong-form-based PINN to approximate the physical solution, which we verify by considering a simpler problem. If the hyperbolic paraboloid is \textit{fully} clamped (i.e., on all four boundaries), no additional terms due to natural boundary conditions enter the loss function of the strong form. Instead, we can incorporate the Dirichlet boundary conditions in $\varphi(\bm\xi)=[(\xi^1)^2-1/4][(\xi^2)^2-1/4]$. Again, we consider $N_{\text{c}}=2{,}048$ collocation points in the reference domain and train the PINN using both the strong and weak form for completeness. As the results in terms of the relative $L_2$-error presented in \figurename~\ref{fig:loss_convergence}b confirm, the PINN using the strong form is indeed able to approximate the FEniCS solution in this simpler setting, although it is still considerably outperformed by the PINN trained via the weak form. For a more intuitive interpretation and comparison of solutions, we plot the results for the fully-clamped case in the physical space in \figurename~\ref{fig:3d_sol_hyperbolic_parab_fc}. This confirms the suitability of the PINN using the strong form in this simpler setting. Nevertheless, the solution is still notably off from the true solution, which is much better approximated by the weak form for the same number of training epochs.

The inability of the PINN to approximate the true solution using the strong form (especially considering the \textit{partly} clamped hyperbolic paraboloid) highlights the inherent challenges of training PINNs with a growing number of terms in the loss function. Although the PINN does have the capability to represent the solution with the given architecture (as observed using the same architecture trained on the weak form), it may not be trivial to identify the corresponding parameters with the standard solvers offered in contemporary ML libraries.

While further strategies to alleviate the training have been proposed, e.g., via scaling the different terms of the loss function so that their gradients are of equal magnitude \cite{Wang2021} or using insights from treating PINNs as so-called Neural Tangent Kernels \cite{Wang2022}, the superiority of the weak form is obvious. Besides drastically higher converge rates, it is also computationally significantly less demanding---both in terms of required operations and storage due to the reduced order of derivatives required. While the training of the strong form took approx.\ $70$s/epoch, using the weak form drastically reduced the training time to approx.\ $3.8$s/epoch ($4.5$s/epoch for $N_{\text{c}}=16{,}384$) on a single Nvidia Quadro RTX 6000. It is for these reasons that we restrict the remaining studies to the weak form only.

\begin{figure}
    \begin{center}
        \import{figures/hyperb_parab/3d_sol}{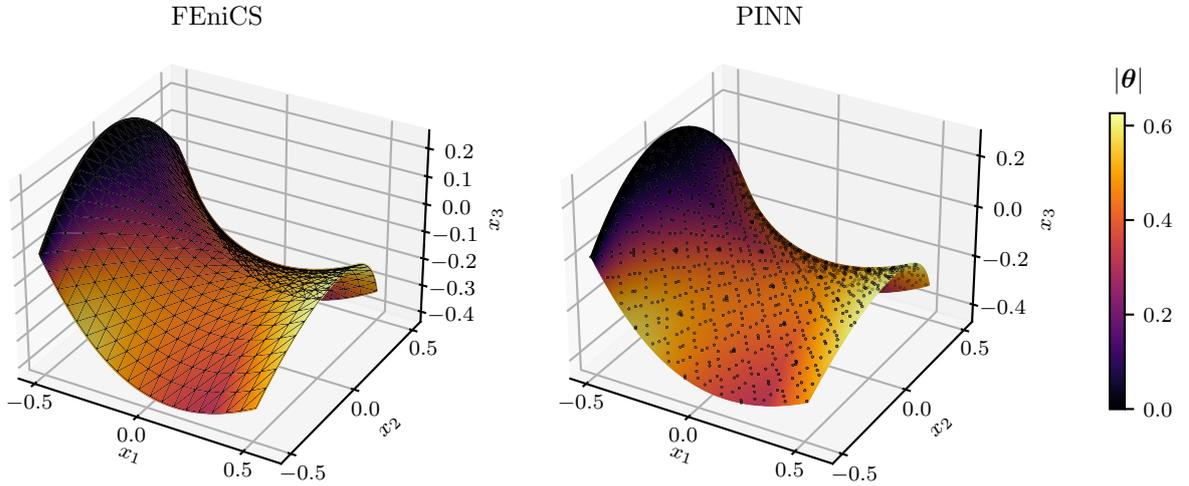}
    \end{center}
    \vspace{-6mm}
    \caption{Comparison of the deformation predicted by the FE and PINN framework (based on the weak form with $N_{\text{c}}=16{,}384$) for the \textit{partly} clamped hyperbolic paraboloid subject to gravity load in the physical space, shown in its deformed configuration. The surface color corresponds to the norm of the rotation fields, $|\bm{\theta}|=|\theta_1|+|\theta_2|$. To highlight the difference in the underlying methods, we display a coarsened version of the selected FE mesh and the first $1{,}024$ points of the selected Sobol sequence in the corresponding plots. Displacements are scaled by a factor of $0.005$ for improved visibility.}\label{fig:3d_sol_hyperb_parab}
\end{figure}

The FEM results and PINN solution based on the weak form (for $N_{\text{c}}=2{,}048$) for the original problem of the \textit{partly} clamped hyperbolic paraboloid are compared in \figurename~\ref{fig:3d_sol_hyperb_parab} (with more detailed information on the relative $L_2$-errors provided in Table~\ref{table:L2_errors}). We observe accurate predictions of the displacement fields throughout the simulation domain, which contains non-trivial membrane- and bending-dominated zones \cite{Hale2018}. This example hence provides a first promising demonstration of the capabilities of PINNs for non-trivial shell mappings. In the following, we continue by analyzing the performance of PINNs for a different, popular shell map, with special emphasis on the thin-thickness limit.

\subsection{Scordelis-Lo roof}
\label{scordelis_lo}
\begin{figure}
    \begin{center}
        \import{figures/scordelis_lo/3d_sol}{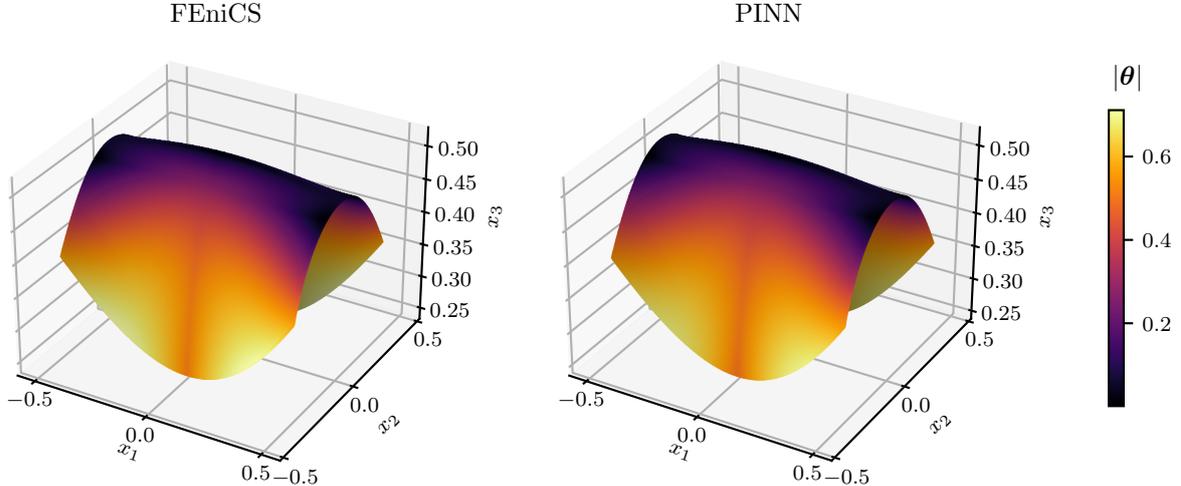}
    \end{center}
    \vspace{-6mm}
    \caption{Comparison of the deformation predicted by the FE and PINN framework (based on the weak form) for the Scordelis-Lo roof benchmark in the physical space, shown in the deformation configuration. The surface color corresponds to the norm of the rotation fields, $|\bm{\theta}|=|\theta_1|+|\theta_2|$. Displacements are scaled by a factor of $0.001$ for improved visibility.} \label{fig:3d_sol_scordelis_lo}
\end{figure}

\begin{figure}
    \begin{center}
        \import{figures/scordelis_lo/sol}{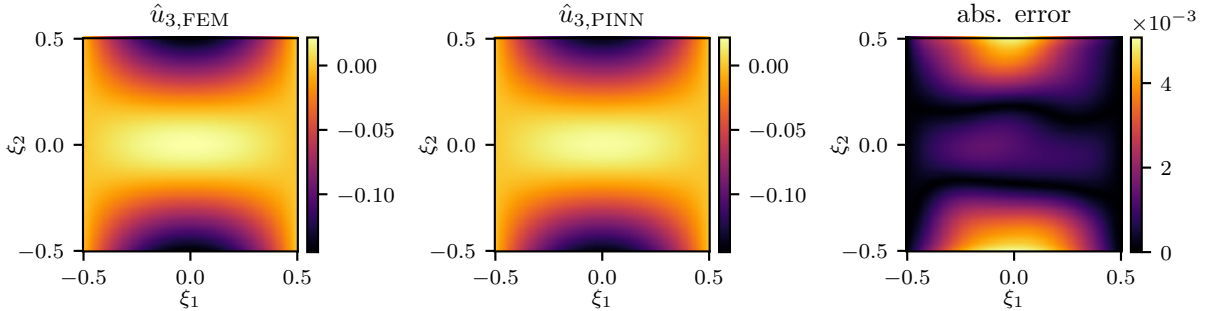}
    \end{center}
    \caption{Comparison of the $\hat{u}_3$-displacement fields predicted by the FE and PINN framework (based on the weak form) for the Scordelis-Lo roof benchmark in the reference domain. Displacements are scaled by a factor of $0.001$ for consistency with Figure~\ref{fig:3d_sol_scordelis_lo}.}\label{fig:u3_scordelis_lo}
\end{figure}

In the second case study, we consider the so-called \textit{Scordelis-Lo roof} benchmark, which is widely used in classical FE settings to test numerical shell implementations against (membrane) locking \cite{Ding2021,Wallner2018,Kefal2020}. The literature on the underlying reasons and methods to mitigate locking is rich and has been of interest for decades, so any attempt at summarizing previous efforts in a few sentences must fall short. We, therefore, refer to the summary given in \cite{Chapelle2011} for in-depth information and restrict to a high-level explanation. Locking arises when the chosen function space of an FE framework is unable to represent the deformation modes of the physical solution. This is especially critical for small shell thicknesses, as pure membrane or bending states, in this case, are typically more pronounced and require a sufficient function space to converge to the correct solution. A variety of methods and improved shell elements have been proposed to mitigate locking in FE setting (see \cite{Chapelle2011} and the references therein). However, their implementation is generally non-trivial and may further rely on parameters that must numerically be estimated for good performance, such as the so-called \textit{splitting parameter} in \cite{Hale2018}. Mathematically robust shell FEs in any membrane- and bending-dominated regime are still out of reach \cite{Hale2018}.

In the following, we numerically investigate the performance of the proposed PINN in a setting where locking is classically observed. We hypothesize that such degeneracies of classical methods should not occur in the PINN setting due to the solution space stemming from the highly nonlinear nature of the multi-layer perceptrons coupled to the selected nonlinear activation function (GELU).

Following the benchmark proposed in \cite{Belytschko1985} and scaling it appropriately, we consider the chart $\bm{\phi}(\xi^1,\xi^2)=\{\xi^1,1/2\,\sin(\xi^2),1/2\,\cos(\xi^2)\}$ with $\xi^1 \in [-1/2,1/2]$ and $\xi^2 \in [-2\pi/9,2\pi/9]$ (corresponding to a cylinder segment of angle $\theta_0=40^{\circ}$, as indicated in \figurename~\ref{fig:case_studies}b, and $L=1$, $r=1/2$), a characteristic thickness of $t/L=0.005$ and $\nu=0$. The shell is subject to gravity loading, so that again\footnote{We here scale the applied gravitational force with $t$ to ensure that the deformation fields do not diverge in the thin-thickness limit. Note that, as before, different force magnitudes simply re-scale the solution fields.} $W_{\text{ext}}=-\int_{\omega}t^2\hat{u}_3 \dd \calS$. Furthermore, the two ends of the cylinder are partly clamped, i.e., $\hat{u}_2=\hat{u}_3=0$ for all $\xi^1=\{-1/2,1/2\}$, so that we may define the trial function $\varphi(\bm\xi)=(\xi^1+1/2)(\xi^1-1/2)$ to account for the Dirichlet boundary conditions. Note that there is a rigid body mode associated with $\hat{u}_1$, which we suppress by fixing the corresponding degree of freedom for a single node in the FE setting and by defining a second trial function $\varphi'(\bm\xi)=(\xi^1)^2+(\xi^2)^2$ to multiply $\hat{u}^*_1$ in the PINN setting.

To evaluate the performance of an implementation, the vertical displacement $\hat{u}_3$ at the midpoint of the free edge is typically compared to the reference solution $\hat{u}_3=-0.3024$ \cite{Belytschko1985}. In addition, we obtain a reference FEM solution from FEniCS simulations (using partial selective reduced integration to mitigate potential locking, as proposed in \cite{Hale2018}). Note that in the original formulation \cite{Belytschko1985} $E=\num{4.32e8}$ and a gravity load of $f=360$ per domain area were chosen, and the physical shell was larger by a factor of 50, so that we must appropriately scale the displacements to compare them with the PINN solution. The solution admits a field partly dominated by pure membrane strains and a boundary layer of pure bending strains, which is why it is selected as a typical test to check for locking. We trained the PINN using the weak form for $1{,}000$ epochs with $N_c=65{,}536$ collocation points. Results are presented in \figurename~\ref{fig:3d_sol_scordelis_lo} and \figurename~\ref{fig:u3_scordelis_lo} (with additional plots for the remaining solution fields presented in \ref{appdx_scordelis}). Remarkably, the PINN is indeed able to approximate the solution fields of the FEM to high accuracy, with the corresponding $L_2$-errors presented in Table~\ref{table:L2_errors}. Both FEM and PINN solutions approach the reference value of \cite{Belytschko1985} closely with $\hat{u}_{3,\text{FEM}}=-0.302$ and $\hat{u}_{3,\text{PINN}}=-0.297$ (if re-scaled consistently). This confirms our hypothesis that PINNs can indeed overcome some of the intricacies of conventional mesh-based methods in the context of shell structures.

\begin{figure}[h]
\centering
\begin{tikzpicture}
\begin{axis}[
    label style={font=\large},
    xlabel={Epoch},
    ylabel={Average rel.\ $L_2$-error},
    xmin=1, xmax=20000,
    ymin=0.03, ymax=1.1,
    xmode=log,
    ymode=log,
    legend pos=outer north east,
    legend cell align=left,
    ymajorgrids=true,
    grid style=dashed,
]
\addplot[color=blue]
    table [x index=0, y index=1, col sep=comma] {figures/plots/thin_thickness_convergence/L2_error_1.csv};

\addplot[color=red]
    table [x index=0, y index=1, col sep=comma] {figures/plots/thin_thickness_convergence/L2_error_01.csv};

\addplot[color=green]
    table [x index=0, y index=1, col sep=comma] {figures/plots/thin_thickness_convergence/L2_error_001.csv}; 

\addplot[color=orange]
    table [x index=0, y index=1, col sep=comma] {figures/plots/thin_thickness_convergence/L2_error_0001.csv}; 

\addplot[color=violet]
    table [x index=0, y index=1, col sep=comma] {figures/plots/thin_thickness_convergence/L2_error_00001_red.csv}; 

\legend{$t/L=1$, $t/L=0.1$, $t/L=0.01$, $t/L=0.001$, $t/L=0.0001$}
  
\end{axis}
\end{tikzpicture}
\caption{\label{fig:convergence_study} Average relative $L_2$-error for the trained epochs of the five solution fields as compared to the FEM solution for different characteristic thicknesses of the Scordelis-Lo roof benchmark. Besides the different ratios of $t/L$, we select the same problem setup and hyperparameters as described in Section~\ref{scordelis_lo}.}
\end{figure}
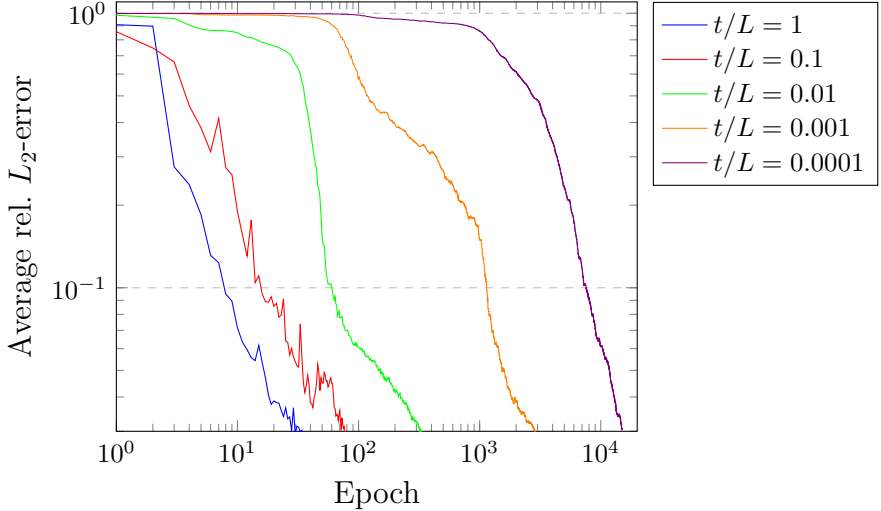

Let us also address the commonly observed convergence problems for small shell thicknesses. The different energy contributions in the weak form, i.e., the membrane, shear, and bending energies, scale differently with thickness $t$. While the former two scale linearly with $t$, the bending energy is proportional to $t^3$, which drastically increases the difference in magnitude of these terms for small values of $t$, especially so for a random initialization of the PINN. In the thin-thickness limit, the corresponding condition number of the Hessian of the loss function with respect to the PINN parameters becomes large, which greatly and negatively impacts the training convergence due to stiff gradient flow dynamics \cite{Wang2021}. A numerical study on the convergence behavior of the proposed PINN for different thickness-to-length ratios is presented in \figurename~\ref{fig:convergence_study}. Training times indeed correlate adversely with the thickness-to-length ratio and may take up to $10{,}000$ epochs to reach an acceptable error for extremely thin shells.

A more quantitative assessment of the impact of ill-posed loss functions on the convergence can be obtained through the lens of the neural tangent kernel \cite{Wang2022}, which has led to ideas on how to artificially scale the different contributions in the loss function to accelerate convergence. While such techniques show promise for loss terms that represent fundamentally different contributions (such as the PDE residual and boundary conditions), it is not possible to leverage such ideas in this setup, since changing the contribution of the different energies also changes the actual physical setup of the shell. To further improve the convergence rate in this setting indeed constitutes a fundamentally different challenge, which is outside of the scope of this work. Nevertheless, the study of the limitations of PINNs under such extreme settings deserves further theoretical and numerical studies, which might also provide a deeper understanding of the performance and limitations of PINNs in more general settings.

\subsection{Hemisphere under concentrated load}

In the last case study we consider a fully clamped hemisphere subject to a vertical load centered at its midpoint. The chart is given by $\bm{\phi}(\xi^1,\xi^2)=\{\xi^1,\xi^2,\sqrt{1-(\xi^1)^2-(\xi^2)^2}\}$ with $(\xi^1,\xi^2) \in D_1$ ($D_1=\{\bfk\in\Rset^2:\|\bfk\|\leq1\}$ denoting the closed unit disc in 2D, and thus $r=1$) and we set $t/r=0.05$ and $\nu=0.3$. The load, approximating a point load, is described by a Gaussian kernel centered at the midpoint as $f(\bm\xi)=\exp[-([\xi^1]^2+[\xi^2])^2/0.1]$ (so that $W_{\text{ext}}=-\int_{\omega}t f(\bm\xi)\hat{u}_3 \dd \calS$), and the clamping is formalized by $\hat{\bm{u}}=\bm{0}$ and $\bm{\theta}=\bm{0}$ for all $(\xi^1,\xi^2) \in S_1$ (with $S_1=\{\bfk\in\Rset^2:\|\bfk\|=1\}$ denoting the unit circle). The clamped boundary conditions are encoded in the trial function $\varphi(\bm\xi)=[1-(\xi^1)^2-(\xi^2)^2]/2$. This elliptical geometry is again fundamentally different from the two previous benchmarks, which were hyperbolic and parabolic, respectively. We here trained the PINN using the weak form for $100$ epochs with $N_c=78{,}400$ collocation points transformed to the unit disc by the low distortion map proposed in \cite{Shirley1997}.

\begin{figure}
    \begin{center}
        \import{figures/hemisphere/3d_sol}{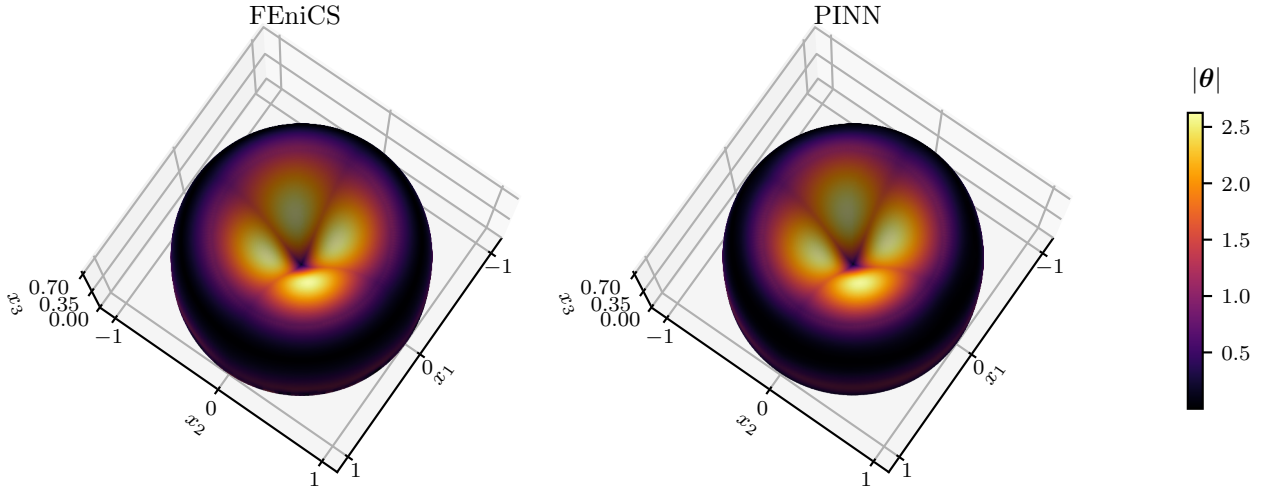}
    \end{center}
    \vspace{-6mm}
    \caption{Comparison of the deformation predicted by the FE and PINN framework (based on the weak form) for the fully clamped hemisphere subject to a concentrated load in the physical space, shown in the deformed configuration. The surface color corresponds to the norm of the rotation fields, $|\bm{\theta}|=|\theta_1|+|\theta_2|$. Displacements are scaled by a factor of $0.05$ for improved visibility.}\label{fig:3d_sol_hemisphere}
\end{figure}

\begin{figure}
    \begin{center}
        \import{figures/hemisphere/sol}{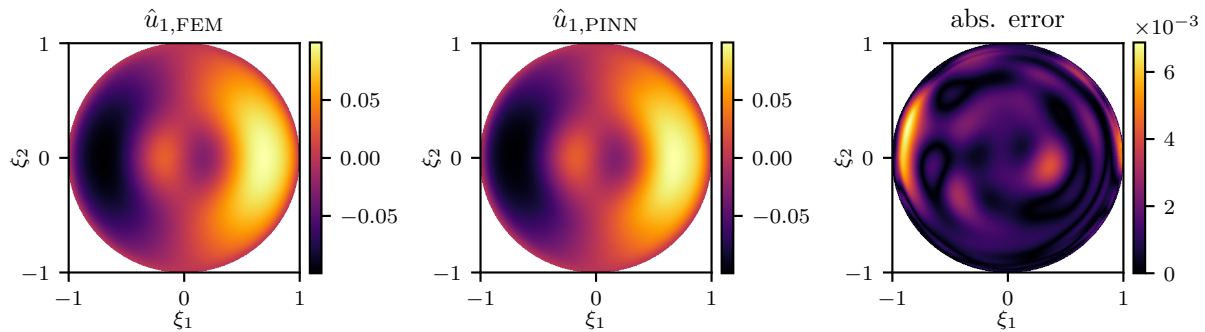}
    \end{center}
    \caption{Comparison of the predicted $\hat{u}_1$-displacement field of the FE and PINN framework (based on the weak form) for the fully clamped hemisphere subject to a concentrated load in the reference domain, shown in the deformed configuration. Displacements are scaled by a factor of $0.05$ for consistency with Figure~\ref{fig:3d_sol_hemisphere}. }\label{fig:u1_hemisphere}
\end{figure}

Results are presented in \figurename~\ref{fig:3d_sol_hemisphere} and \figurename~\ref{fig:u1_hemisphere}, with additional plots for the remaining solution fields available in \ref{appdx_hemisphere}. Again, the solution obtained with the PINN compares well to the FEM solution; the $L_2$-errors are summarized in Table~\ref{table:L2_errors}, indicating a high degree of agreement. 

\subsection{Discussion}

The three benchmarks outlined above indicate a level of robustness of PINNs for solving variational problems on manifolds, as the PINNs perform well regardless of the given chart and corresponding curvature as well as for different reference domains, boundary and loading conditions. Using the weak form has clear advantages over the strong form and generally leads to a good agreement between PINN predictions and FEM-based results. While PINNs may not (yet) compete with traditional solvers in terms of computational efficiency in the small-thicknesses limit, it is worth highlighting that we did not observe any fundamental failure of PINNs to approximate the true solution fields, even in the thin-thickness limit, and therefore can indeed numerically confirm the absence of locking in the evaluated scenarios. Also, note that the presented (shearable) Naghdi shell formulation is typically applied to rather thick shells, as we can otherwise neglect the shear contributions and consider a more suitable, simpler (unshearable) shell theory. While the different thickness scaling between the membrane and bending energies and the resulting ill-posed loss function in the thin-thickness limit will remain, such simplified models will likely facilitate the training to some extent.

\section{Conclusion}
\label{sec:Conclusions}

We have presented a physically-informed neural network (PINN) approach to solve for the mechanical (small-strain) response of arbitrarily shaped shell structures. Using Naghdi shell theory, the formulation is based on three translational and two rotational degrees of freedom. Since this theory is defined on the midsurface manifold of the shell, the PINN must solve the shell equations in a non-Euclidean domain. Results for three classical benchmark examples demonstrate that PINNs are indeed capable of solving the shell equations on arbitrary manifolds and that results compare well to solutions obtained via FEM.

In agreement with previous studies, considering the strong form of the equations as the PINN loss leads to poor convergence in certain settings, mainly due to the increased amount of (competing) loss terms compared to the weak form, which consistently outperforms the strong form. Furthermore, we observe worse convergence with decreasing shell thickness, which we attribute to stronger changes in the solution fields and, more importantly, the increasingly imbalanced loss function.
Nevertheless, we demonstrated that PINNs show potential in representing the physical solution in this thin-thickness limit, for which classical methods are prone to show locking---without applying numerical adaptations that are usually not straightforward to implement. On the contrary, implementing the shell equations in the PINN setting is comparably elegant and simple, and it does not require any adjustments to successfully predict solutions in the small-thickness setting, as demonstrated by successfully passing the Scordelis-Lo roof benchmark. One can therefore exploit PINN-based methods such as the one presented here to serve as a sanity check for FEM to ensure locking-free behavior.

It would be of interest to further study the convergence behavior of PINNs in the small-thickness limit towards an improved understanding of the challenges of training PINNs in a general sense. Besides, further improvements to reduce the computational efforts are necessary to make PINNs competitive with well-established FEM techniques. Lastly, their straightforward extension to inverse problems or topology optimization is promising, as the computational efficiency may quickly surpass the FEM setting and autodifferentiation provides analytical sensitivities for free.

\subsection*{Data and code availability}
The PINN framework used in the current study is publicly available at [link to be inserted].

\subsection*{Acknowledgements}
The authors thank Raidas Rimkus for his support in establishing the Naghdi shell formulation and implementation.

\bibliography{MyCollection}

\begin{thebibliography}{10}
\expandafter\ifx\csname url\endcsname\relax
  \def\url#1{\texttt{#1}}\fi
\expandafter\ifx\csname urlprefix\endcsname\relax\def\urlprefix{URL }\fi
\expandafter\ifx\csname href\endcsname\relax
  \def\href#1#2{#2} \def\path#1{#1}\fi

\bibitem{Naghdi1973}
P.~M. Naghdi, \href{http://link.springer.com/10.1007/978-3-662-39776-3_5}{{The
  Theory of Shells and Plates}}, in: Linear Theories of Elasticity and
  Thermoelasticity, Springer Berlin Heidelberg, Berlin, Heidelberg, 1973, pp.
  425--640.
\newblock \href {https://doi.org/10.1007/978-3-662-39776-3_5}
  {\path{doi:10.1007/978-3-662-39776-3_5}}.
\newline\urlprefix\url{http://link.springer.com/10.1007/978-3-662-39776-3_5}

\bibitem{Simulia2014}
Simulia, \href{http://130.149.89.49:2080/v6.14/books/usb/default.htm}{{Abaqus
  Analysis User's Guide}}, Tech. rep. (2014).
\newline\urlprefix\url{http://130.149.89.49:2080/v6.14/books/usb/default.htm}

\bibitem{COMSOL2019}
COMSOL, {COMSOL Multiphysics Reference Manual}, Tech. rep. (2019).

\bibitem{Hale2018}
J.~S. Hale, M.~Brunetti, S.~P. Bordas, C.~Maurini,
  \href{https://doi.org/10.1016/j.compstruc.2018.08.001}{{Simple and extensible
  plate and shell finite element models through automatic code generation
  tools}}, Computers and Structures 209 (2018) 163--181.
\newblock \href {https://doi.org/10.1016/j.compstruc.2018.08.001}
  {\path{doi:10.1016/j.compstruc.2018.08.001}}.
\newline\urlprefix\url{https://doi.org/10.1016/j.compstruc.2018.08.001}

\bibitem{Hale2012}
J.~Hale, P.~Baiz,
  \href{https://linkinghub.elsevier.com/retrieve/pii/S0045782512001983}{{A
  locking-free meshfree method for the simulation of shear-deformable plates
  based on a mixed variational formulation}}, Computer Methods in Applied
  Mechanics and Engineering 241-244 (2012) 311--322.
\newblock \href {https://doi.org/10.1016/j.cma.2012.06.010}
  {\path{doi:10.1016/j.cma.2012.06.010}}.
\newline\urlprefix\url{https://linkinghub.elsevier.com/retrieve/pii/S0045782512001983}

\bibitem{Chapelle2011}
D.~Chapelle, K.-J. Bathe,
  \href{http://link.springer.com/10.1007/978-3-642-16408-8}{{The Finite Element
  Analysis of Shells - Fundamentals}}, Computational Fluid and Solid Mechanics,
  Springer Berlin Heidelberg, Berlin, Heidelberg, 2011.
\newblock \href {https://doi.org/10.1007/978-3-642-16408-8}
  {\path{doi:10.1007/978-3-642-16408-8}}.
\newline\urlprefix\url{http://link.springer.com/10.1007/978-3-642-16408-8}

\bibitem{Lecun2015}
Y.~Lecun, Y.~Bengio, G.~Hinton, {Deep learning}, Nature 521~(7553) (2015)
  436--444.
\newblock \href {https://doi.org/10.1038/nature14539}
  {\path{doi:10.1038/nature14539}}.

\bibitem{Zheng2021}
L.~Zheng, S.~Kumar, D.~M. Kochmann,
  \href{https://linkinghub.elsevier.com/retrieve/pii/S0045782521002310}{{Data-driven
  topology optimization of spinodoid metamaterials with seamlessly tunable
  anisotropy}}, Computer Methods in Applied Mechanics and Engineering 383
  (2021) 113894.
\newblock \href {https://doi.org/10.1016/j.cma.2021.113894}
  {\path{doi:10.1016/j.cma.2021.113894}}.
\newline\urlprefix\url{https://linkinghub.elsevier.com/retrieve/pii/S0045782521002310}

\bibitem{Bastek2022}
J.-H. Bastek, S.~Kumar, B.~Telgen, R.~N. Glaesener, D.~M. Kochmann,
  \href{https://pnas.org/doi/full/10.1073/pnas.2111505119}{{Inverting the
  structure–property map of truss metamaterials by deep learning}},
  Proceedings of the National Academy of Sciences 119~(1) (jan 2022).
\newblock \href {https://doi.org/10.1073/pnas.2111505119}
  {\path{doi:10.1073/pnas.2111505119}}.
\newline\urlprefix\url{https://pnas.org/doi/full/10.1073/pnas.2111505119}

\bibitem{Lagaris1998}
I.~Lagaris, A.~Likas, D.~Fotiadis,
  \href{http://ieeexplore.ieee.org/document/712178/}{{Artificial neural
  networks for solving ordinary and partial differential equations}}, IEEE
  Transactions on Neural Networks 9~(5) (1998) 987--1000.
\newblock \href {https://doi.org/10.1109/72.712178}
  {\path{doi:10.1109/72.712178}}.
\newline\urlprefix\url{http://ieeexplore.ieee.org/document/712178/}

\bibitem{Raissi2019}
M.~Raissi, P.~Perdikaris, G.~Karniadakis,
  \href{https://linkinghub.elsevier.com/retrieve/pii/S0021999118307125}{{Physics-informed
  neural networks: A deep learning framework for solving forward and inverse
  problems involving nonlinear partial differential equations}}, Journal of
  Computational Physics 378 (2019) 686--707.
\newblock \href {https://doi.org/10.1016/j.jcp.2018.10.045}
  {\path{doi:10.1016/j.jcp.2018.10.045}}.
\newline\urlprefix\url{https://linkinghub.elsevier.com/retrieve/pii/S0021999118307125}

\bibitem{Cai2022}
S.~Cai, Z.~Mao, Z.~Wang, M.~Yin, G.~E. Karniadakis,
  \href{https://link.springer.com/10.1007/s10409-021-01148-1}{{Physics-informed
  neural networks (PINNs) for fluid mechanics: a review}}, Acta Mechanica
  Sinica (jan 2022).
\newblock \href {https://doi.org/10.1007/s10409-021-01148-1}
  {\path{doi:10.1007/s10409-021-01148-1}}.
\newline\urlprefix\url{https://link.springer.com/10.1007/s10409-021-01148-1}

\bibitem{Cai2021}
S.~Cai, Z.~Wang, S.~Wang, P.~Perdikaris, G.~E. Karniadakis,
  \href{https://asmedigitalcollection.asme.org/heattransfer/article/143/6/060801/1104439/Physics-Informed-Neural-Networks-for-Heat-Transfer}{{Physics-Informed
  Neural Networks for Heat Transfer Problems}}, Journal of Heat Transfer
  143~(6) (jun 2021).
\newblock \href {https://doi.org/10.1115/1.4050542}
  {\path{doi:10.1115/1.4050542}}.
\newline\urlprefix\url{https://asmedigitalcollection.asme.org/heattransfer/article/143/6/060801/1104439/Physics-Informed-Neural-Networks-for-Heat-Transfer}

\bibitem{Haghighat2021}
E.~Haghighat, M.~Raissi, A.~Moure, H.~Gomez, R.~Juanes,
  \href{https://linkinghub.elsevier.com/retrieve/pii/S0045782521000773}{{A
  physics-informed deep learning framework for inversion and surrogate modeling
  in solid mechanics}}, Computer Methods in Applied Mechanics and Engineering
  379 (2021) 113741.
\newblock \href {https://doi.org/10.1016/j.cma.2021.113741}
  {\path{doi:10.1016/j.cma.2021.113741}}.
\newline\urlprefix\url{https://linkinghub.elsevier.com/retrieve/pii/S0045782521000773}

\bibitem{Mishra2022}
S.~Mishra, R.~Molinaro,
  \href{https://academic.oup.com/imajna/advance-article/doi/10.1093/imanum/drab093/6503953}{{Estimates
  on the generalization error of physics-informed neural networks for
  approximating PDEs}}, IMA Journal of Numerical Analysis (jan 2022).
\newblock \href {https://doi.org/10.1093/imanum/drab093}
  {\path{doi:10.1093/imanum/drab093}}.
\newline\urlprefix\url{https://academic.oup.com/imajna/advance-article/doi/10.1093/imanum/drab093/6503953}

\bibitem{Wang2022}
S.~Wang, X.~Yu, P.~Perdikaris,
  \href{https://linkinghub.elsevier.com/retrieve/pii/S002199912100663X}{{When
  and why PINNs fail to train: A neural tangent kernel perspective}}, Journal
  of Computational Physics 449 (2022) 110768.
\newblock \href {https://doi.org/10.1016/j.jcp.2021.110768}
  {\path{doi:10.1016/j.jcp.2021.110768}}.
\newline\urlprefix\url{https://linkinghub.elsevier.com/retrieve/pii/S002199912100663X}

\bibitem{Li2021}
W.~Li, M.~Z. Bazant, J.~Zhu,
  \href{https://linkinghub.elsevier.com/retrieve/pii/S004578252100270X}{{A
  physics-guided neural network framework for elastic plates: Comparison of
  governing equations-based and energy-based approaches}}, Computer Methods in
  Applied Mechanics and Engineering 383 (2021) 113933.
\newblock \href {https://doi.org/10.1016/j.cma.2021.113933}
  {\path{doi:10.1016/j.cma.2021.113933}}.
\newline\urlprefix\url{https://linkinghub.elsevier.com/retrieve/pii/S004578252100270X}

\bibitem{Zhuang2021}
X.~Zhuang, H.~Guo, N.~Alajlan, H.~Zhu, T.~Rabczuk,
  \href{https://doi.org/10.1016/j.euromechsol.2021.104225
  https://linkinghub.elsevier.com/retrieve/pii/S099775382100019X}{{Deep
  autoencoder based energy method for the bending, vibration, and buckling
  analysis of Kirchhoff plates with transfer learning}}, European Journal of
  Mechanics - A/Solids 87~(November 2020) (2021) 104225.
\newblock \href {https://doi.org/10.1016/j.euromechsol.2021.104225}
  {\path{doi:10.1016/j.euromechsol.2021.104225}}.
\newline\urlprefix\url{https://doi.org/10.1016/j.euromechsol.2021.104225
  https://linkinghub.elsevier.com/retrieve/pii/S099775382100019X}

\bibitem{Yan2022}
C.~Yan, R.~Vescovini, L.~Dozio,
  \href{https://doi.org/10.1016/j.compstruc.2022.106761}{{A framework based on
  physics-informed neural networks and extreme learning for the analysis of
  composite structures}}, Computers \& Structures 265 (2022) 106761.
\newblock \href {https://doi.org/10.1016/j.compstruc.2022.106761}
  {\path{doi:10.1016/j.compstruc.2022.106761}}.
\newline\urlprefix\url{https://doi.org/10.1016/j.compstruc.2022.106761}

\bibitem{Tang2021}
Z.~Tang, Z.~Fu, \href{http://arxiv.org/abs/2103.02811}{{Physics-informed Neural
  Networks for Elliptic Partial Differential Equations on 3D Manifolds}} (mar
  2021).
\newblock \href {http://arxiv.org/abs/2103.02811} {\path{arXiv:2103.02811}}.
\newline\urlprefix\url{http://arxiv.org/abs/2103.02811}

\bibitem{Gaile2011}
S.~Gaile, {Free material optimization for shells and plates} (2011).

\bibitem{Ciarlet2005}
P.~G. Ciarlet, \href{http://link.springer.com/10.1007/s10659-005-4738-8}{{An
  Introduction to Differential Geometry with Applications to Elasticity}},
  Journal of Elasticity 78-79~(1-3) (2005) 1--215.
\newblock \href {https://doi.org/10.1007/s10659-005-4738-8}
  {\path{doi:10.1007/s10659-005-4738-8}}.
\newline\urlprefix\url{http://link.springer.com/10.1007/s10659-005-4738-8}

\bibitem{Chapelle1998}
D.~Chapelle, R.~Stenberg,
  \href{http://epubs.siam.org/doi/10.1137/S0036142996302918}{{Stabilized Finite
  Element Formulations for Shells in a Bending Dominated State}}, SIAM Journal
  on Numerical Analysis 36~(1) (1998) 32--73.
\newblock \href {https://doi.org/10.1137/S0036142996302918}
  {\path{doi:10.1137/S0036142996302918}}.
\newline\urlprefix\url{http://epubs.siam.org/doi/10.1137/S0036142996302918}

\bibitem{Kingma2015}
D.~P. Kingma, J.~Ba, \href{http://arxiv.org/abs/1412.6980}{{Adam: A Method for
  Stochastic Optimization}}, 3rd International Conference on Learning
  Representations, ICLR 2015 - Conference Track Proceedings (2014) 1--15\href
  {http://arxiv.org/abs/1412.6980} {\path{arXiv:1412.6980}}.
\newline\urlprefix\url{http://arxiv.org/abs/1412.6980}

\bibitem{Liu1989}
D.~C. Liu, J.~Nocedal, \href{http://link.springer.com/10.1007/BF01589116}{{On
  the limited memory BFGS method for large scale optimization}}, Mathematical
  Programming 45~(1-3) (1989) 503--528.
\newblock \href {https://doi.org/10.1007/BF01589116}
  {\path{doi:10.1007/BF01589116}}.
\newline\urlprefix\url{http://link.springer.com/10.1007/BF01589116}

\bibitem{E2018}
W.~E, B.~Yu, \href{http://link.springer.com/10.1007/s40304-018-0127-z}{{The
  Deep Ritz Method: A Deep Learning-Based Numerical Algorithm for Solving
  Variational Problems}}, Communications in Mathematics and Statistics 6~(1)
  (2018) 1--12.
\newblock \href {https://doi.org/10.1007/s40304-018-0127-z}
  {\path{doi:10.1007/s40304-018-0127-z}}.
\newline\urlprefix\url{http://link.springer.com/10.1007/s40304-018-0127-z}

\bibitem{Sobol1967}
I.~Sobol',
  \href{https://linkinghub.elsevier.com/retrieve/pii/0041555367901449}{{On the
  distribution of points in a cube and the approximate evaluation of
  integrals}}, USSR Computational Mathematics and Mathematical Physics 7~(4)
  (1967) 86--112.
\newblock \href {https://doi.org/10.1016/0041-5553(67)90144-9}
  {\path{doi:10.1016/0041-5553(67)90144-9}}.
\newline\urlprefix\url{https://linkinghub.elsevier.com/retrieve/pii/0041555367901449}

\bibitem{Sukumar2022}
N.~Sukumar, A.~Srivastava,
  \href{https://linkinghub.elsevier.com/retrieve/pii/S0045782521006186}{{Exact
  imposition of boundary conditions with distance functions in physics-informed
  deep neural networks}}, Computer Methods in Applied Mechanics and Engineering
  389 (2022) 114333.
\newblock \href {https://doi.org/10.1016/j.cma.2021.114333}
  {\path{doi:10.1016/j.cma.2021.114333}}.
\newline\urlprefix\url{https://linkinghub.elsevier.com/retrieve/pii/S0045782521006186}

\bibitem{Jagtap2020}
A.~D. Jagtap, K.~Kawaguchi, G.~E. Karniadakis,
  \href{https://linkinghub.elsevier.com/retrieve/pii/S0021999119308411}{{Adaptive
  activation functions accelerate convergence in deep and physics-informed
  neural networks}}, Journal of Computational Physics 404 (2020) 109136.
\newblock \href {https://doi.org/10.1016/j.jcp.2019.109136}
  {\path{doi:10.1016/j.jcp.2019.109136}}.
\newline\urlprefix\url{https://linkinghub.elsevier.com/retrieve/pii/S0021999119308411}

\bibitem{Wight2020}
C.~L. Wight, J.~Zhao, \href{http://arxiv.org/abs/2007.04542}{{Solving
  Allen-Cahn and Cahn-Hilliard Equations using the Adaptive Physics Informed
  Neural Networks}} (jul 2020).
\newblock \href {http://arxiv.org/abs/2007.04542} {\path{arXiv:2007.04542}}.
\newline\urlprefix\url{http://arxiv.org/abs/2007.04542}

\bibitem{Krishnapriyan2021}
A.~S. Krishnapriyan, A.~Gholami, S.~Zhe, R.~M. Kirby, M.~W. Mahoney,
  \href{http://arxiv.org/abs/2109.01050}{{Characterizing possible failure modes
  in physics-informed neural networks}} (sep 2021).
\newblock \href {http://arxiv.org/abs/2109.01050} {\path{arXiv:2109.01050}}.
\newline\urlprefix\url{http://arxiv.org/abs/2109.01050}

\bibitem{Hendrycks2016}
D.~Hendrycks, K.~Gimpel, \href{http://arxiv.org/abs/1606.08415}{{Gaussian Error
  Linear Units (GELUs)}} (jun 2016).
\newblock \href {http://arxiv.org/abs/1606.08415} {\path{arXiv:1606.08415}}.
\newline\urlprefix\url{http://arxiv.org/abs/1606.08415}

\bibitem{Paszke2019}
A.~Paszke, S.~Gross, F.~Massa, A.~Lerer, J.~Bradbury, G.~Chanan, T.~Killeen,
  Z.~Lin, N.~Gimelshein, L.~Antiga, A.~Desmaison, A.~K{\"{o}}pf, E.~Yang,
  Z.~DeVito, M.~Raison, A.~Tejani, S.~Chilamkurthy, B.~Steiner, L.~Fang,
  J.~Bai, S.~Chintala, \href{http://arxiv.org/abs/1912.01703}{{PyTorch: An
  imperative style, high-performance deep learning library}}, Advances in
  Neural Information Processing Systems (dec 2019).
\newblock \href {http://arxiv.org/abs/1912.01703} {\path{arXiv:1912.01703}}.
\newline\urlprefix\url{http://arxiv.org/abs/1912.01703}

\bibitem{Belytschko1985}
T.~Belytschko, H.~Stolarski, W.~K. Liu, N.~Carpenter, J.~S. Ong,
  \href{https://linkinghub.elsevier.com/retrieve/pii/0045782585900350}{{Stress
  projection for membrane and shear locking in shell finite elements}},
  Computer Methods in Applied Mechanics and Engineering 51~(1-3) (1985)
  221--258.
\newblock \href {https://doi.org/10.1016/0045-7825(85)90035-0}
  {\path{doi:10.1016/0045-7825(85)90035-0}}.
\newline\urlprefix\url{https://linkinghub.elsevier.com/retrieve/pii/0045782585900350}

\bibitem{Wang2021}
S.~Wang, Y.~Teng, P.~Perdikaris,
  \href{https://epubs.siam.org/doi/10.1137/20M1318043}{{Understanding and
  Mitigating Gradient Flow Pathologies in Physics-Informed Neural Networks}},
  SIAM Journal on Scientific Computing 43~(5) (2021) A3055--A3081.
\newblock \href {https://doi.org/10.1137/20M1318043}
  {\path{doi:10.1137/20M1318043}}.
\newline\urlprefix\url{https://epubs.siam.org/doi/10.1137/20M1318043}

\bibitem{Ding2021}
C.~Ding, K.~K. Tamma, H.~Lian, Y.~Ding, T.~J. Dodwell, S.~P.~A. Bordas,
  \href{https://link.springer.com/10.1007/s00466-020-01944-9}{{Uncertainty
  quantification of spatially uncorrelated loads with a reduced-order
  stochastic isogeometric method}}, Computational Mechanics 67~(5) (2021)
  1255--1271.
\newblock \href {https://doi.org/10.1007/s00466-020-01944-9}
  {\path{doi:10.1007/s00466-020-01944-9}}.
\newline\urlprefix\url{https://link.springer.com/10.1007/s00466-020-01944-9}

\bibitem{Wallner2018}
M.~Wallner, C.~Birk, H.~Gravenkamp,
  \href{https://onlinelibrary.wiley.com/doi/10.1002/pamm.201800381}{{Numerical
  analysis of thin‐walled structures based on the scaled boundary finite
  element method}}, PAMM 18~(1) (dec 2018).
\newblock \href {https://doi.org/10.1002/pamm.201800381}
  {\path{doi:10.1002/pamm.201800381}}.
\newline\urlprefix\url{https://onlinelibrary.wiley.com/doi/10.1002/pamm.201800381}

\bibitem{Kefal2020}
A.~Kefal, E.~Oterkus,
  \href{https://www.mdpi.com/1424-8220/20/9/2685}{{Isogeometric iFEM Analysis
  of Thin Shell Structures}}, Sensors 20~(9) (2020) 2685.
\newblock \href {https://doi.org/10.3390/s20092685}
  {\path{doi:10.3390/s20092685}}.
\newline\urlprefix\url{https://www.mdpi.com/1424-8220/20/9/2685}

\bibitem{Shirley1997}
P.~Shirley, K.~Chiu,
  \href{http://www.tandfonline.com/doi/abs/10.1080/10867651.1997.10487479}{{A
  Low Distortion Map Between Disk and Square}}, Journal of Graphics Tools 2~(3)
  (1997) 45--52.
\newblock \href {https://doi.org/10.1080/10867651.1997.10487479}
  {\path{doi:10.1080/10867651.1997.10487479}}.
\newline\urlprefix\url{http://www.tandfonline.com/doi/abs/10.1080/10867651.1997.10487479}

\end{thebibliography}

\newpage
\appendix


\section{Derivation of the strain measures}
\label{app:Derivation}

We here provide the mathematical steps required in the derivation of the Green-Lagrange strain tensor components. Starting from \eqref{eq:linear_green_lagrange}, i.e.,
\be\label{eq:linear_green_lagrange2}
    E_{ij} = \frac{1}{2} \big(\bm{g}_i \bm{U}_{,j} + \bm{g}_j \bm{U}_{,i} \big),
\ee
we compute the above derivatives as (omitting the dependency on $\xi^{i}$ for conciseness)
\be
\frac{\partial \bm{U}}{\partial \xi^{\alpha}} = \frac{\partial \bm{u}}{\partial \xi^{\alpha}} + \xi^3 \frac{\partial (\theta_{\lambda}\bm{a}^{\,\lambda})}{\partial \xi^{\alpha}}.
\ee
Note that these partial derivatives are taken along a curve on the shell's midsurface with changing base vectors. Therefore, evaluation of the above derivatives leads to
\be
\begin{split}
\frac{\partial \bm{u}}{\partial \xi^{\alpha}} &= 
\frac{\partial( u_{\lambda} \bm{a}^{\,\lambda} + u_3 \bm{a}^{\, 3})} {\partial \xi^{\alpha}} = u_{\lambda| \alpha} \bm{a}^{\, \lambda} + b^{\lambda}_{\alpha} u_{\lambda} \bm{a}_3 +  u_{3,\lambda} \bm{a}^{\, 3} - u_3 b_{\lambda \alpha} \bm{a}^{\, \lambda},\\
\frac{\partial (\theta_{\lambda} \bm{a}^{\, \lambda})}{\partial \xi^{\alpha}} &=  \theta_{\lambda| \alpha} \bm{a}^{\, \lambda} + b^{\lambda}_{\alpha} \theta_{\lambda} \bm{a}_3,
\end{split}
\ee
and 
\be
\frac{\partial \bm{U}}{\partial \xi^{3}} = \theta_{\lambda} \bm{a}^{\, \lambda}.
\ee
If we further replace the 3D basis vectors $\bm{g}_i$ in \eqref{eq:linear_green_lagrange2} by \eqref{eq:basis_transformation}, we find the Green-Lagrange strain tensor components as
\begin{align} 
\begin{split}
E_{\alpha \beta} &= \frac{1}{2} ( u_{\alpha | \beta} + u_{\beta | \alpha} ) - b_{\alpha \beta} u_3 \\[4pt]
 &+\xi^3 \Bigg[\frac{1}{2} \bigg( \theta_{\alpha | \beta} + \theta_{\beta | \alpha} - b^{\lambda}_{\beta} u_{\lambda | \alpha } - b^{\lambda}_{\alpha} u_{\lambda | \beta } \bigg) + c_{\alpha \beta} u_3\Bigg]\\[4pt]
 &+ (\xi^3)^2 \Bigg[ \frac{1}{2} \bigg( b^{\lambda}_{\beta} \theta_{\lambda | \alpha } + b^{\lambda}_{\alpha} \theta_{\lambda | \beta } \bigg)   \Bigg],
\end{split}\\[4pt]
E_{\alpha 3} &= \frac{1}{2}(\theta_{\alpha} + u_{3, \alpha} + b^{\lambda}_{\alpha} u_{\lambda}),\\[4pt]
E_{33} &= 0.
\end{align}

\section{Solution details for the presented benchmarks}

In this section, we summarize the solution fields not shown in the main article for all three benchmarks along with the $L_2$-errors for each case and solution field.

\subsection{Solution fields for the partly clamped hyperbolic paraboloid}\label{appdx_hyperb}
\begin{figure}[H]
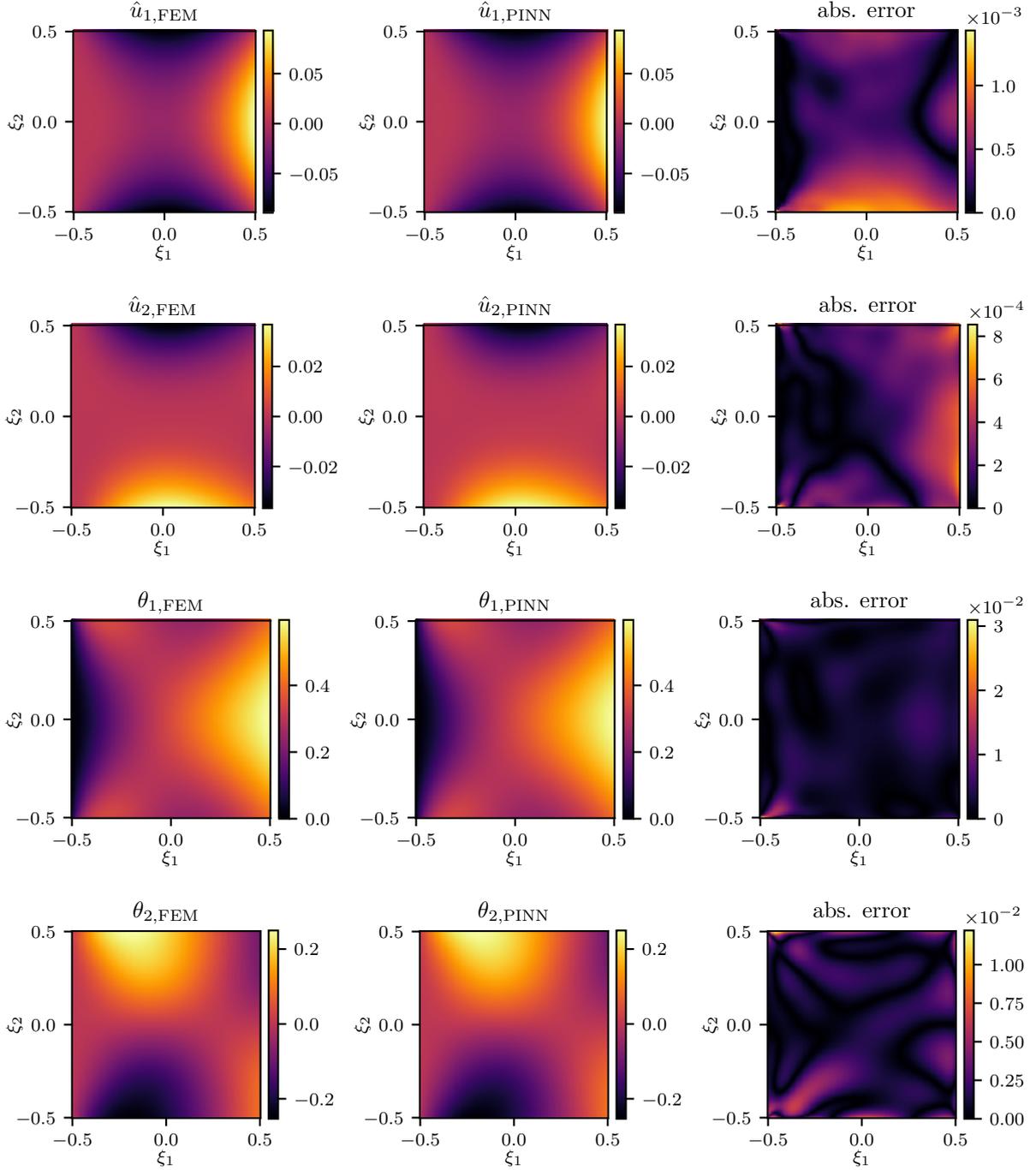

    \begin{center}
        \import{figures/hyperb_parab/sol_4}{fig0.pgf}
        \import{figures/hyperb_parab/sol_4}{fig1.pgf}
        \import{figures/hyperb_parab/sol_4}{fig3.pgf}
        \import{figures/hyperb_parab/sol_4}{fig4.pgf}
    \end{center}
    \caption{Comparison of the $\hat{u}_1,\hat{u}_2,\theta_1,\theta_2$-solution fields predicted by the FE and PINN framework (based on the weak form with $N_{\text{c}}=16{,}384$) for the partly clamped hyperbolic paraboloid subject to gravity load in the reference domain. Displacements are scaled by a factor of $0.005$ for consistency with Figure~\ref{fig:3d_sol_hyperb_parab}.}
    \label{fig:hyperb_parab_sol4}
\end{figure}

\subsection{Solution fields for the Scordelis-Lo roof}
\enlargethispage{5\baselineskip}
\label{appdx_scordelis}
\begin{figure}[H]
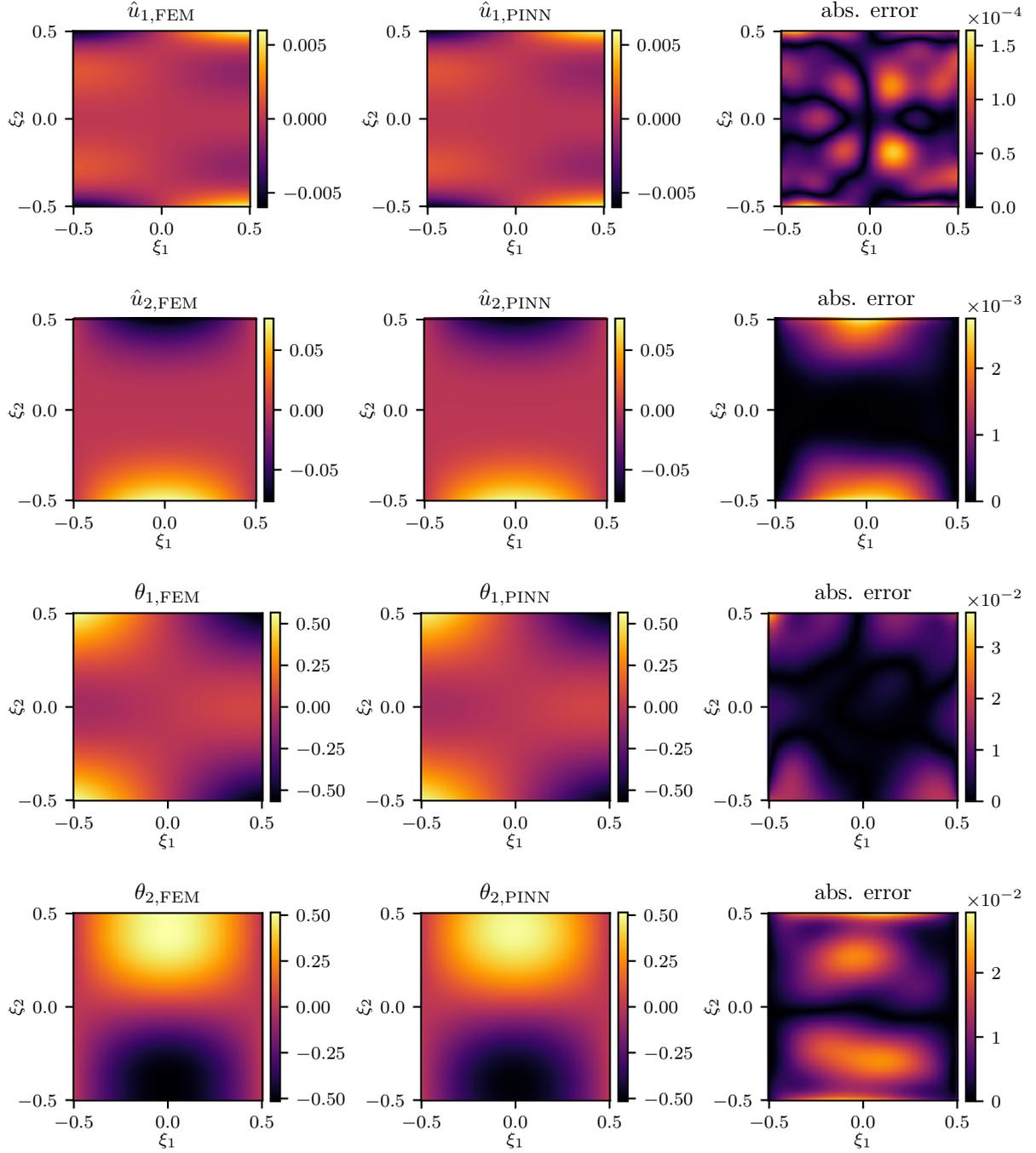

    \begin{center}
        \import{figures/scordelis_lo/sol_4}{fig0.pgf}
        \import{figures/scordelis_lo/sol_4}{fig1.pgf}
        \import{figures/scordelis_lo/sol_4}{fig3.pgf}
        \import{figures/scordelis_lo/sol_4}{fig4.pgf}
    \end{center}
    \caption{Comparison of the $\hat{u}_1,\hat{u}_2,\theta_1,\theta_2$-solution fields predicted by the FE and PINN framework (based on the weak form) for the Scordelis-Lo roof benchmark in the reference domain. Displacements are scaled by a factor of $0.001$ for consistency with Figure~\ref{fig:3d_sol_scordelis_lo}.}
    \label{fig:scordelis_lo_sol4}
\end{figure}

\vspace*{-4mm}

\subsection{Solution fields for the fully clamped hemisphere}
\label{appdx_hemisphere}
\begin{figure}[H]
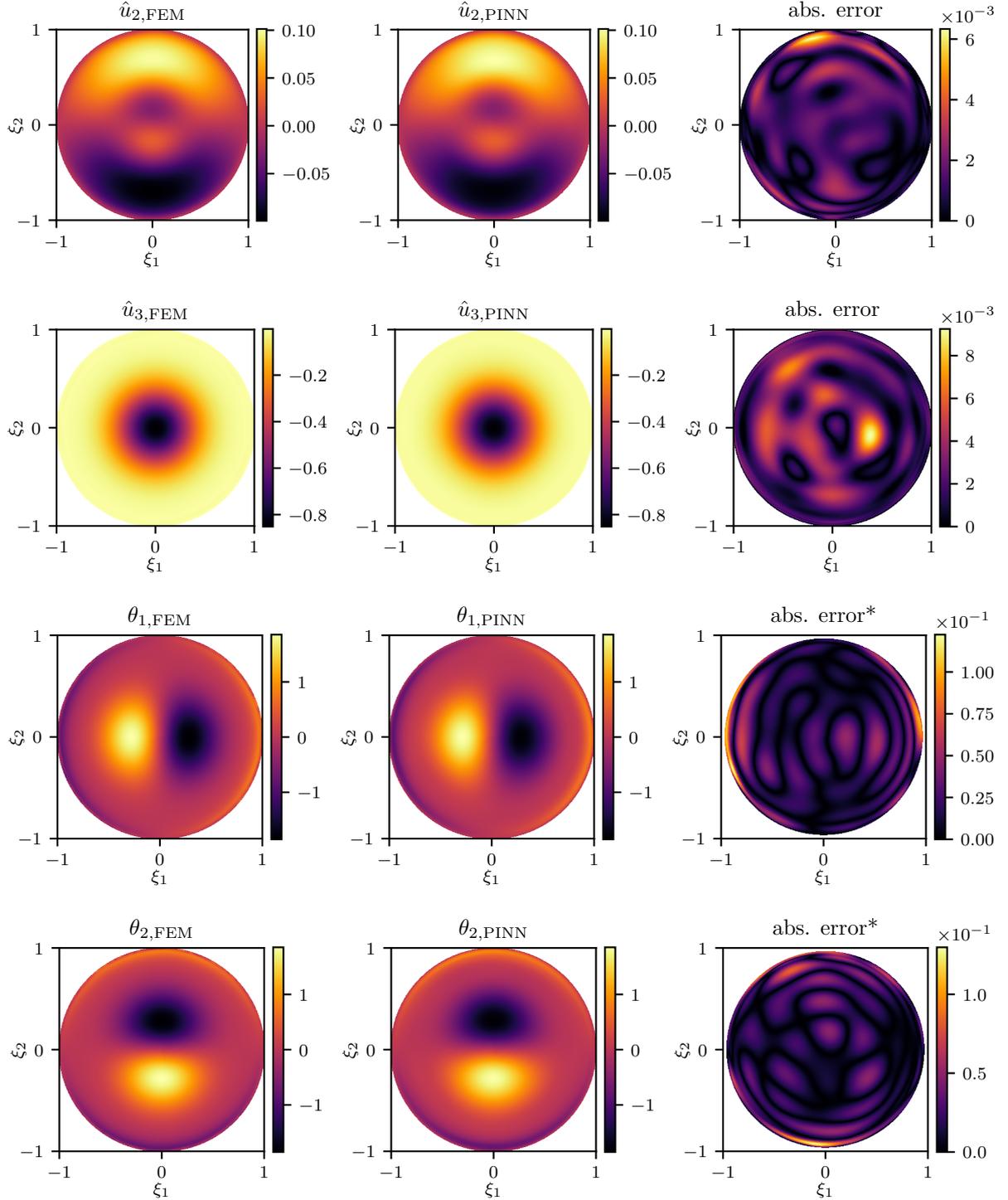

    \begin{center}
        \import{figures/hemisphere/sol_4}{fig1.pgf}
        \import{figures/hemisphere/sol_4}{fig2.pgf}
        \import{figures/hemisphere/sol_4}{fig3.pgf}
        \import{figures/hemisphere/sol_4}{fig4.pgf}
    \end{center}
    \vspace{-5mm}
    \caption{Comparison of the $\hat{u}_2,\hat{u}_3,\theta_1,\theta_2$-solution fields predicted by the FE and PINN framework (based on the weak form) for the fully clamped hemisphere subject to a concentrated load in the reference domain. Displacements are scaled by a factor of $0.05$ for consistency with Figure~\ref{fig:3d_sol_hemisphere}. $^*$We only consider a radius up to $0.96$, as otherwise artifacts from the FE mesh dominate the error plots, see also Figure~\ref{fig:hemisphere_mesh}.}
    \label{fig:hemisphere_sol4}
\end{figure}

\begin{figure}
	\centering
    \includegraphics[width=0.7\textwidth]{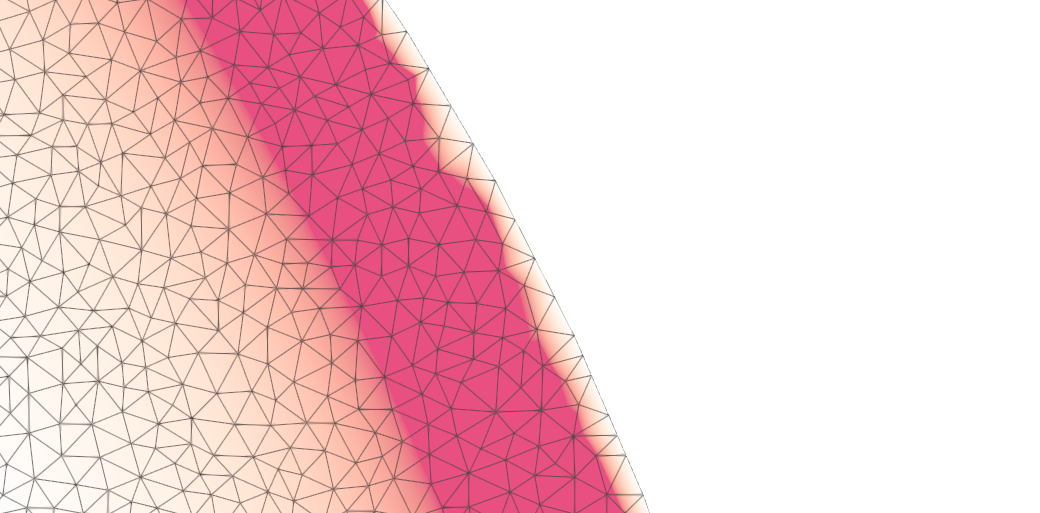}
	\caption{Unphysical artifacts in the $\theta_{1}$ solution field of the FE framework due to non-uniform meshing.}
	\label{fig:hemisphere_mesh}
\end{figure}

Note that we observed artifacts in the $\theta_{1,2}$ solution fields close to the boundary in the FEM solution, which is due to mesh irregularities, as quantitatively shown for $\theta_{1}$ in Figure~\ref{fig:hemisphere_mesh}. Further studies showed that this effect persists even with increased mesh resolution and is indeed an inherent consequence of the mesh partition. While a fully regular mesh might mitigate this effect, this is only attainable for regular geometries and highlights another advantage of the mesh-free PINN.

\subsection{Errors of the three reported benchmarks}

\begin{table}[!h]
\begin{center}
     \caption{Relative $L_2$-error of the PINN solution fields based on the weak form, compared to the FEM solutions for the three considered case studies (including in parentheses in (a) the errors for the increased collocation point sampling with $N_{\text{c}}=16{,}384$).}
     \begin{tabular}[t]{cc}
     \multicolumn{2}{c}{(a) Hyperbolic paraboloid}\\
      \hline
      $\hat{u}_1$ & $0.107$ ($0.0139$) \\
      $\hat{u}_2$ & $0.104$ ($0.017$) \\
      $\hat{u}_3$ & $0.0502$ ($0.00833$) \\
      $\theta_1$ & $0.0426$ ($0.00759$) \\
      $\theta_2$ & $0.0652$ ($0.0125$) \\ 
    \end{tabular}
    \hspace{1em}
     \begin{tabular}[t]{cc}
     \multicolumn{2}{c}{(b) Scordelis-Lo roof} \\
      \hline
      $\hat{u}_1$ & $0.0417$ \\
      $\hat{u}_2$ & $0.0362$ \\
      $\hat{u}_3$ & $0.0353$ \\
      $\theta_1$ & $0.0353$ \\
      $\theta_2$ & $0.0372$ \\ 
    \end{tabular}
    \hspace{1em}
     \begin{tabular}[t]{cc}
     \multicolumn{2}{c}{(c) Hemisphere} \\
      \hline
      $\hat{u}_1$ & $0.0287$ \\
      $\hat{u}_2$ & $0.0274$ \\
      $\hat{u}_3$ & $0.00648$ \\
      $\theta_1$ & $0.0329$ \\
      $\theta_2$ & $0.0277$ \\ 
    \end{tabular}
    \label{table:L2_errors}
\end{center}
\end{table}


\end{document}